\documentclass{ws-jnopm}

\usepackage{graphics}
\usepackage{graphicx}
\usepackage{amsfonts}
\usepackage{amssymb}
\usepackage{amsmath}
\usepackage{multirow}
\usepackage{color}
\usepackage{chngcntr}
\usepackage{array}
\usepackage[hypertex, colorlinks, colorlinks=false, linkcolor=blue, citecolor=green, filecolor=black, urlcolor=cyan]{hyperref}

\begin{document}

\markboth{Rick Lytel, Sean M. Mossman, Mark Kuzyk}{Optimization of eigenstates \ldots}

\catchline{}{}{}{}{}

\title{OPTIMIZATION OF EIGENSTATES AND SPECTRA FOR QUASI-LINEAR NONLINEAR OPTICAL SYSTEMS}

\author{RICK LYTEL, SEAN M. MOSSMAN, and MARK G. KUZYK}

\address{Department of Physics and Astronomy, Washington State University\\ Pullman, Washington 99164-2814\\
rlytel@wsu.edu}

\maketitle


\begin{history}
\end{history}

\begin{abstract}
Quasi-one-dimensional quantum structures with spectra scaling faster than the square of the eigenmode number (superscaling) can generate intrinsic, off-resonant optical nonlinearities near the fundamental physical limits, independent of the details of the potential energy along the structure. The scaling of spectra is determined by the topology of the structure, while the magnitudes of the transition moments are set by the geometry of the structure. This paper presents a comprehensive study of the geometrical optimization of superscaling quasi-one-dimensional structures and provides heuristics for designing molecules to maximize intrinsic response.  A main result is that designers of conjugated structures should attach short side groups at least a third of the way along the bridge, not near its end as is conventionally done.  A second result is that once a side group is properly placed, additional side groups do not further enhance the response.
\end{abstract}


\maketitle

\section{Introduction}\label{sec:intro}
Nonlinear optics is the study of quantum systems with polarizabilities that are nonlinear functions of external electromagnetic fields.  Nonlinear optical phenomena include harmonic generation \cite{maker65.02,bloem68.01,bass69.01}, optical phase conjugation\cite{yariv77.01,yariv78.01} and four-wave mixing\cite{boyd92.01,lytel86.02}, optical bistability\cite{winfu80.01,gibbs84.01}, ultrafast optics \cite{weine11.01}, and waveguide switching \cite{lytel84.01,vanec91.01}.  Materials with suitable polarizabilities have been sought in solids\cite{baugh78.06,lipsc81.01,hugga87.01,cheml80.01}, liquids\cite{giord65.01,giord67.01,ho79.01,chen88.01}, liquid crystals\cite{barni83.01,chen88.01}, and gases\cite{shelt82.01,kaatz98.01}, as well as in photonic crystals, mesoscopic solid state wires\cite{guble99.01,foste08.01,tian09.01}, and artificial systems\cite{cheml85.01,miller1984band,rink89.01,schmi87.01}.  Applications center on electric-field control of light for ultrafast optical signal modulation using the third-rank optical susceptibility tensor $\chi_{ijk}$ and all-optical, light by light control for high-speed fiber-optic switching using the fourth rank optical susceptibility tensor $\chi_{ijkl}$.  Ultrafast effects, which rely on virtual transitions among electronic states, arise from the far-from-resonance response to an electric field.  A prime objective of research in nonlinear optics is the discovery of the fundamental origins of large intrinsic responses that may be realized in macroscopic materials and artificial structures.

Fundamental quantum mechanics limits the intrinsic nonlinearities of any molecular structure\cite{kuzyk00.01,kuzyk09.01,kuzyk06.03,kuzyk03.01}.  Until 2006, a factor of thirty gap existed between theoretical and experimental values of the first hyperpolarizability\cite{kuzyk13.01}.  The existence of the gap suggested that conventional molecular design rules required modification to enhance the intrinsic nonlinear optical response.  Advances in the design and synthesis of conjugated chain polyynes\cite{slepk04.01,eisle05.01,stefk13.01}, and extended conjugation and substituted donor-acceptor cyanoethynylethene molecules\cite{may07.01,burev07.01} produced molecules with larger third-order nonlinear optical susceptibilities, as did modulated conjugation molecular structures \cite{zhou06.01,zhou07.02,perez07.01,perez09.01}.

For off-resonant phenomena, the eigenstates and spectra of the system are the critical parameters that control the hyperpolarizabilities. The first comprehensive study of the relation between spectral properties and the nonlinear optical response was a Monte Carlo calculation\cite{shafe11.01} which showed that a necessary condition for a large intrinsic response is that the energy spectrum should scale faster than the square of the eigenmode number, so-called superscaling spectra\cite{lytel14.01}. Superscaling spectra resemble particle-in-the-box energies, with periodic boundaries separating randomly positioned wavenumbers.  It has been shown that the potential in donor-acceptor configurations can produce such spectra.

Donor-acceptor structures with superscaling spectra require another design rule to achieve maximum response, one that specifies the relative distributions of the lowest eigenstates lying above the Fermi level in a multielectron system.  Near the quantum limits, the response for the first hyperpolarizability is dominated by the contributions from only three eigenstates, the so-called three-state conjecture\cite{kuzyk14.01}.  For the second hyperpolarizability, the response is dominated by four eigenstates.  The optimum transition moments are those creating spatial charge separation among the ground and excited states, leading to a large change in the dipole moments $\bar{x}_{nn}\equiv x_{nn}-x_{00}$ for $n=1,2$, and with an additional requirement that the geometry maximize the alignment of the lowest order off-diagonal transition moment contributions $x_{01}, x_{02}, x_{12}$ to the response.

This paper is the third in a series of studies of the optimum topology and geometry for quasi-one-dimensional structures, such as nanowires or donor-acceptor molecules, by analyzing the eigenstates and spectra of one electron quantum graph models with bare and dressed edges.  The first study showed that dressed quantum graphs could achieve nonlinearities near the fundamental limit\cite{lytel13.04}, while the second revealed the optimum topologies\cite{lytel14.01} for producing such spectra.  In this paper, we show that the optimum geometries for systems with side groups and optimum spectra are those that generate the requisite change in dipole moments while maintaining a large spatial overlap among the transition moments between the lowest three or four eigenstates.  The quantum graph models suggest that a general rule for designing real structures is to place the side groups at specific locations along the chains, but \emph{far from the ends}.  Such configurations create eigenfunctions producing large changes in the dipole moments in the presence of an external optical field while at the same time allowing good spatial overlap for large off-diagonal transition moments.

Quantum graphs are quasi-1D systems to which electron dynamics are confined along the edge of a network structure which can be thought of as a branched nano-wire structure, or a quasi-linear molecule, such as a donor-acceptor, with side groups.  Quantum graphs were first studied as tractable molecular models\cite{pauli36.01,kuhn48.01,ruede53.01,scher53.01,platt53.01} and have been invoked as models of mesoscopic systems\cite{kowal90.01}, optical waveguides \cite{flesi87.01}, quantum wires\cite{ivche98.01,sanch98.01}, excitations in fractals \cite{avish92.01}, and fullerines, graphene, and carbon nanotubes\cite{amovi04.01,leys04.01,kuchm07.01}. Quantum graphs are also exactly solvable models of quantum chaos\cite{kotto97.01,kotto99.02,kotto00.01,blume01.04}.  Microwave networks have recently been successfully used to experimentally simulate quantum graphs\cite{hul04.01}.

Quantum graphs were recently introduced into nonlinear optics\cite{shafe12.01,lytel12.01,lytel13.01,lytel13.03,lytel13.04,lytel14.01} as model systems exhibiting the optimum spectra for large intrinsic nonlinear optical response\cite{lytel14.01}, for specific geometries.  In this paper, we exploit quasi-one-dimensional quantum graphs in a comprehensive study of the optimum geometries of nonlinear optical structures with particular focus on the placement of a defect or side group along a main chain such that the spatial distribution of the lowest eigenfunctions produces transition moments that optimize the response.

All of the graphs analyzed in this paper are of the form of a wire supporting a $\delta$ function potential simulating a defect, a side prong simulating a side group, and combinations of these.  Typical donor-acceptor molecules contain two identical groups on each end, one nearly collinear with the main chain and the other at an angle to it.  Dressed and prong wire graphs simulate the geometry of donor-acceptor molecules, but their dynamics are obviously an idealization of a many electron molecular system.  However, the single electron model shows explicitly the form and shape of the states and spectra that real systems must possess.  Since chemists are expert at shaping states and spectra with strategic placement of side groups along main chains, the heuristics described here should prove quite useful to them in their pursuit of molecular designs with the geometry and topology required to bridge the gap.

This paper is organized as follows.  Section \ref{sec:NLOgraphs} briefly reviews the calculation of the first and second hyperpolarizabilities of quantum graphs with arbitrary shape and topology.  Section \ref{sec:delta} discusses graphs dressed with $\delta$ functions.  Section \ref{sec:prong} discusses the graphs with prongs only.  Section \ref{sec:deltaANDprong} discusses dressed graphs with prongs.  We show that $\delta$ graphs with a positive potential are spectrally similar to prong graphs for the lowest energy eigenfunctions on the graphs.  We derive a design heuristic for the graphs to generate the optimum transition moments.  Section \ref{sec:end} presents our conclusions and discusses the applications of the heuristics to molecular design.  A set of appendices reviews two other kinds of graphs, those with square wells or steps in \ref{sec:squareWells}, and those with slant wells in \ref{sec:slantWells}.

\section{Nonlinear optics of planar quantum graphs}\label{sec:NLOgraphs}
The calculation of optical nonlinearities for quantum graphs is articulated in\cite{shafe12.01,lytel13.01}.  In brief, the eigenfunctions and spectra are determined by solving the nonrelativistic wave equation for the edges of the graph and applying conservation conditions at the vertices.  The spectra and transition moments are then used in a sum over states to compute intrinsic second and third order nonlinearities \cite{kuzyk00.01}.  Tensor properties are easily studied using spherical representations of the Cartesian tensors $\beta_{ijk}$ and $\gamma_{ijkl}$, as described in \cite{lytel12.01}. All quantum graphs have superscaling spectra, with wavenumbers contained within root boundaries spaced more or less equally.  The following subsections summary our standard method.

\subsection{A quasi-1D dynamical system}
The dynamics of an electron on a quantum graph are described by a self-adjoint Hamiltonian operating on the edges of the graph, with complex amplitude and probability conservation (hereafter referred to as flux conservation throughout the paper) at all internal vertices and fixed, infinite potentials at the termination vertices (where the amplitude vanishes).  The physics of the eigenfunctions and their spectra have been previously described, along with a suitable lexicography for describing the \emph{union} operation for creating eigenfunctions from the edge functions that solve the equations of motion for the Hamiltonian\cite{shafe12.01}.

\begin{figure}\centering
\includegraphics[width=3.4in]{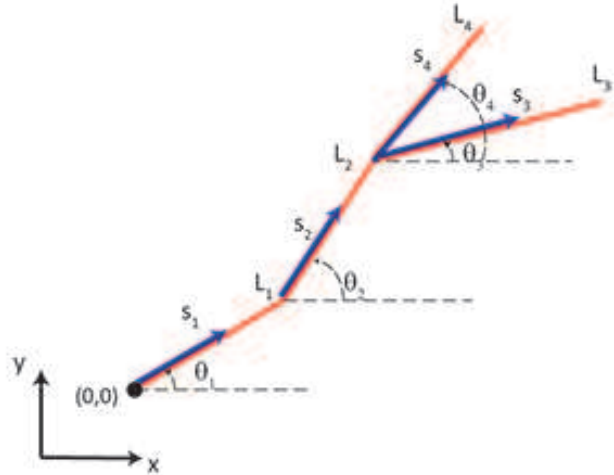}
\caption{A four-edge quantum graph.  Each edge has its own longitudinal coordinate $s_{i}$ ranging from zero to $L_{i}$.  The projection $x(s_{i})$ of an edge onto the $x-axis$ is measured from the origin of the coordinate system attached to the graph (and not to the beginning of the edge itself).  For example, $x(s_1)=s_1\cos\theta_1$ while $x(s_2)=L_1\cos\theta_1 + s_2\cos\theta_2$ and so on.}\label{fig:graphNEW}
\end{figure}

The graph is specified by the location of its vertices and the edges connecting the vertices.  Figure \ref{fig:graphNEW} details the notation and configuration of a graph.  A set of vertices with arbitrary locations in the 2D plane but fixed connections specifies a topological class of graphs.  For a fixed topology, the variation of vertex locations specifies various geometries for the graph.  Since motion is confined to the graph edges and is continuous at each vertex, the energy spectrum depends only on the edge lengths and the boundary conditions, ie, the topology.  Spectra are quasi-quadratic in state number, ie, superscaling.  This is the first requirement for a large response from a nonlinear optical structure.

The edge lengths and angular positions determine the projections of electron motion onto a fixed, external reference axis.  The projections summed over all edges yield the transition moments required to compute the tensor elements of the hyperpolarizability tensors.  Regardless of how the axes used to define the vertices are chosen, the various tensor components may be used to assemble any component in a different frame by using the rotation properties of the tensors.

The study of the nonlinear optical properties of a specific graph topology requires solving the graph for its eigenfunctions and spectra as functions of its edge lengths and using them to compute a set of transition moments for the graph from which the hyperpolarizability tensors may be computed.

\subsection{States and spectra}\label{statesAndSpectra}

A general quantum graph is solved for its spectra and states once the Hamiltonian has been specified.  Then, the Hamiltonian is used to compute a set of edge functions, $\phi_n^{i}(s_{i})$ that are solutions to the Schr\"{o}dinger equation with the same set of eigenvalues $E_{n}$ on each edge, but with specific boundary conditions determined by their connections at the vertices.  Next, the edge functions are used to construct eigenfunctions of the Hamiltonian for the graph through union process that reflects a direct sum Hilbert space over the edges \cite{shafe12.01}:
\begin{equation}\label{eigenFunction}
\psi_n(s)= \cup_{i=1}^{E} \phi_n^{i}(s_{i})
\end{equation}
The union is defined such that the eigenfunction is continuous at every vertex of the graph, while the probability current is conserved at each vertex and thus throughout the graph.  These two boundary conditions guarantee that the eigenfunctions are complete.  They also generate the relationships among the edge amplitudes required to compute the spectrum of the graph.  Solutions of the amplitude equations resulting from the boundary conditions exist only if the determinant of the matrix of coefficients vanishes.  This condition produces the secular or characteristic equation for the graph and determines the eigenvalues $k_{n}$ and the exact energy spectrum $E_{n}$.  Since the boundary conditions in the elementary QG model are independent of the angles the edges make with respect to one another, the secular equation is independent of angles and depends only on dimensionless parameters $k_{n}a_{i}$.  For a given configuration of vertices, the distance between them and the rules by which they are connected, i.e., the topology of the graph, determines the energy spectrum.  Topologically different graphs with identical geometries have different energy spectra.  In this way, the graph topology has a large impact on the nonlinear optical response.

Except for bent wires and closed loops, the secular equation of a graph is generally a transcendental equation.  Accurate solutions are easily found numerically.  From these, the relative internal amplitudes may be calculated.  Normalizing the eigenfunction produces the states required to compute the transition moments.

The transition moments are sums (not unions) over edges of the following form:
\begin{equation}\label{xNM}
x_{nm}=\sum_{i=1}^{E}\int_{0}^{a_{i}}\phi_{n}^{*i}(s_{i})\phi_{m}^{i}(s_{i})\ x(s_{i})ds_{i}
\end{equation}
where $\phi_{m}^{i}(s_{i})$ is the normalized wave function on the $i^{th}$ edge that obeys the boundary conditions for the graph and $x(s_{i})$ is the x-component of $s_{i}$, measured from the origin of the graph (and not of the edge), and is a function of the prior edge lengths and angles.

The transverse wavefunction is not calculated and not essential in this model.  The transverse states do not contribute to the hyperpolarizabilities\cite{shafe11.02}.  But there are residual effects from the transverse state in the sum rules, as has been previously discussed\cite{shafe11.02,shafe12.01}.

The process for calculating the spectra and transition moments of elementary QG's can be summarized as follows: (1)  select a particular graph topology, specifying the number of vertices and the connecting edges, (2)  generate a random set of vertices, and calculate the lengths of the edges and the angles each makes with the $x$-axis of the graph's coordinate system, (3) solve the Schr\"{o}dinger equation on each edge of the graph, and (4) match boundary conditions at the vertices and terminal points.  This results in a set of equations for the amplitudes of the wavefunctions on each edge.  The solvability of this set requires that the determinant of the amplitude coefficients vanishes, leading to a secular equation for the energies.  The transition moments $x_{nm}$ and energies may be used to compute the first and second hyperpolarizabilities of any graph specified by a set of vertices, as described next.

\subsection{Hyperpolarizabilities of planar quantum graphs}\label{subsec:quasiQG}

For the quasi-1D problem the tensors are indexed in the (x,y) directions.  The full tensor expressions are given as a sum over states.  The first intrinsic hyperpolarizability tensor for 2D graphs may then be written as
\begin{eqnarray}\label{betaInt}
\beta_{ijk} &\equiv& \frac{\beta}{\beta_{max}} = \left(\frac{3}{4}\right)^{3/4} {\sum_{n,m}}' \frac{\xi_{0n}^{i}\bar{\xi}_{nm}^{j}\xi_{m0}^{k}}{e_n e_m} ,
\end{eqnarray}
where $\xi_{nm}^{i}$ and $e_n$ are normalized transition moments and energies, defined by
\begin{equation}\label{xNMnorm}
\xi_{nm}^{i} = \frac{r_{nm}^{i}}{r_{01}^{max}}, \qquad e_{n} = \frac{E_{n0}}{E_{10}},
\end{equation}
with $r^{(i=1)}=x$ and $r^{(i=2)}=y$, $E_{nm}=E_n-E_m$, and where
\begin{equation}\label{Xmax}
r_{01}^{max} = \left(\frac{\hbar^2}{2 m E_{10}}\right)^{1/2}.
\end{equation}
$r_{01}^{max}$ represents the largest possible transition moment value of $r_{01}$\cite{kuzyk00.01}.  According to Eq. (\ref{xNMnorm}), $e_0 = 0$ and $e_1 = 1$.  $\beta_{ijk}$ has been normalized to its maximum value, is scale-invariant, and can be used to compare molecules of different shapes and sizes.
Similarly, the second intrinsic hyperpolarizability is given by
\begin{eqnarray}\label{gammaInt}
\gamma_{ijkl} &=& \frac{1}{4} \left({\sum_{n,m,l}}' \frac{\xi_{0n}^{i}\bar{\xi}_{nm}^{j}\bar{\xi}_{ml}^{k}\xi_{l0}^{l}}{e_n e_m e_l} - {\sum_{n,m}}' \frac{\xi_{0n}^{i}\xi_{n0}^{j}\xi_{0m}^{j}\xi_{m0}^{k}}{e_n^2 e_m}\right) . \nonumber \\
\end{eqnarray}
The maximum values used for normalization are derived from the Thomas-Reiche-Kuhn sum rules\cite{kuzyk00.01,kuzyk09.01,kuzyk06.03,kuzyk03.01}, and are
\begin{equation}\label{sh-betaMax}
\beta_{max} = 3^{1/4} \left(\frac{e\hbar}{m^{1/2}}\right)^3 \frac{N^{3/2}}{E_{10}^{7/2}}
\end{equation}
and
\begin{equation}\label{sh-gammaMax}
\gamma_{max} = 4 \left(\frac{e^4\hbar^4}{m^2}\right) \frac{N^{2}}{E_{10}^{5}} .
\end{equation}
where $E_{10}$ is the ground state to first excited state energy gap, and $N$ is the number of electrons.  The second hyperpolarizability normalized this way has a largest negative value equal to $-(1/4)$ of the maximum value.  Hyperpolarizabilities normalized this way enable direct comparisons of the intrinsic response without regard to size.

The fundamental limits for $\beta_{xxx}$ fall short of unity for every system calculated from a potential energy function\cite{shafe13.01,lytel13.04,ather12.01} and appear to asymptote to a value of $0.7089$.  For $\gamma_{xxxx}$, the apparent maximum and minimum limits are sixty percent of the values predicted by Eq. (\ref{sh-gammaMax}).  For the rest of this paper, we use the term \emph{fundamental limits} to mean the \emph{apparent} fundamental limits.

\section{Dressed graphs}\label{sec:delta}

\begin{figure}\center
\includegraphics[width=2.5in]{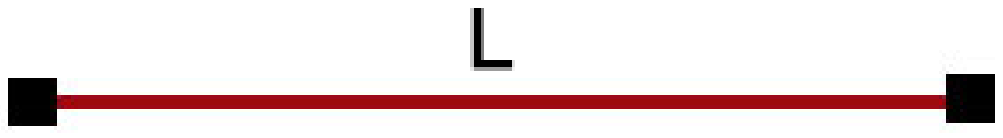}\includegraphics[width=2.5in]{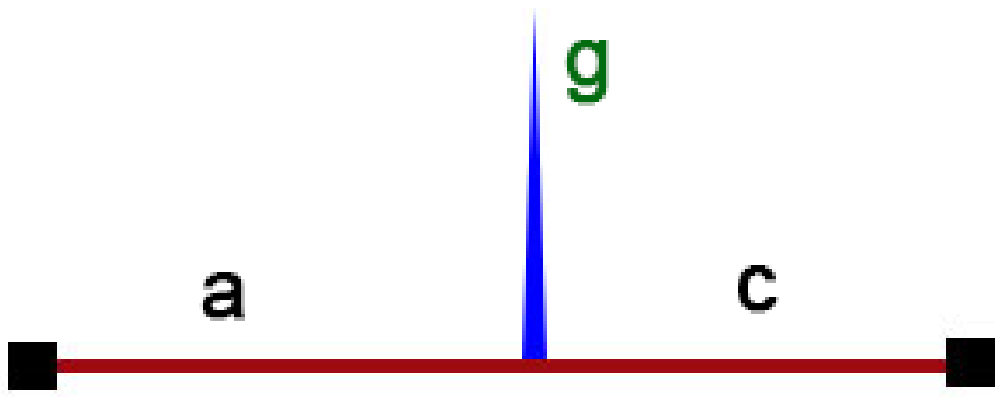}\\
\includegraphics[width=2.5in]{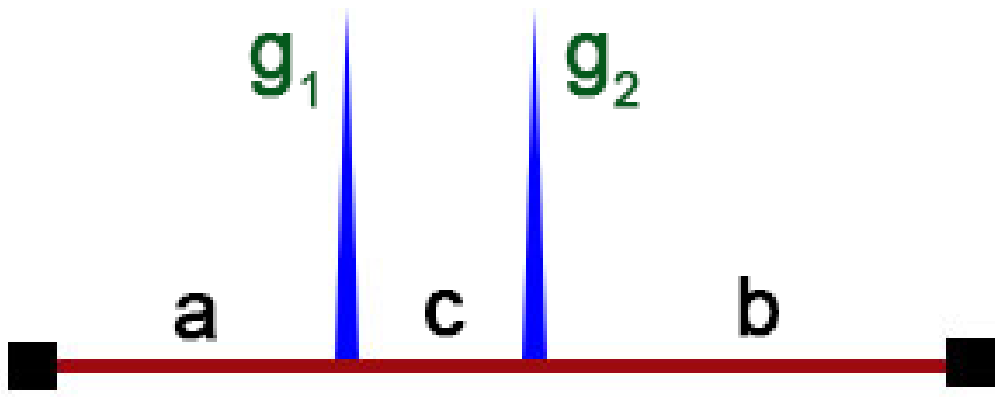}\includegraphics[width=2.5in]{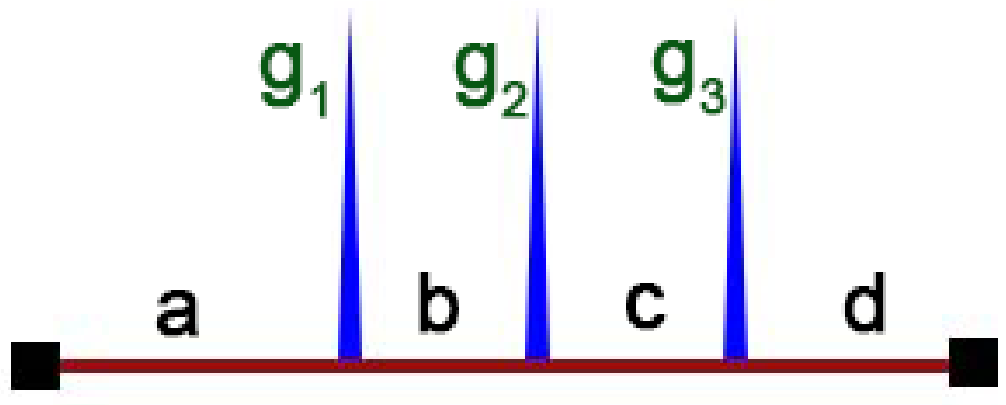}\\
\caption{The bare wire graph and variants with one, two, and three $\delta$ potentials, respectively.  The probability flux at the position of each $\delta$ potential is discontinuous by an amount determined by the strength of the potential.}
\label{fig:DeltaWireGraphs}
\end{figure}

Figure \ref{fig:DeltaWireGraphs} shows a bare wire graph and three variants, each with a modification of the bare wire to introduce finite $\delta$ potentials. The addition of one or more $\delta$ functions to a wire creates a source or sink of probability flux\cite{lytel13.04}.  This has a profound effect on the spectrum at the lower end, and on the shape of the ground state.  Both effects are expected to cause an enhancement in the hyperpolarizabilities.  The details of how this comes about are described in this section.

\subsection{Bare wire}\label{bareWire}

Figure \ref{fig:DeltaWireGraphs}, upper left panel, shows a bare wire quantum graph consisting of an edge supporting electron dynamics, bounded at either end by an infinite potential.  The electron behaves as a particle-in-a-box, with $\beta_{xxx}=0$, and $\gamma_{xxxx}=-0.126$.  However, the graph is a quasi-1D object, and if the wire is bent or is comprised of two edges with a nonzero angle between them, $\beta_{xxx}$ is nonzero, and $\gamma_{xxxx}$ changes slightly.  The first seven eigenfunctions for this graph are shown in the left panel of Figure \ref{fig:1deltaWireZeroBetaBestBetaSide2side}.  For the bare wire (one edge), the centrosymmetry keeps $\beta_{xxx}$ at zero.  The probability flux in the well varies from one edge to the other in a continuous, conserved fashion.  The probability that the electron is on the edge is always unity.

We now examine variations of this graph that have been modified to include finite $\delta$ potentials.  The modifications create interesting quantum graph structures that are straightforward to solve, have more complicated energy spectra, and have eigenfunctions whose shapes are altered in such a way as to create giant enhancements of the optical nonlinearities by localizing the ground state to the $\delta$ potentials, while pushing the first few excited states out from the ground state.

\subsection{Wire with one $\delta$ potential}\label{1deltaWire}

\begin{figure}\center
\includegraphics[width=5in]{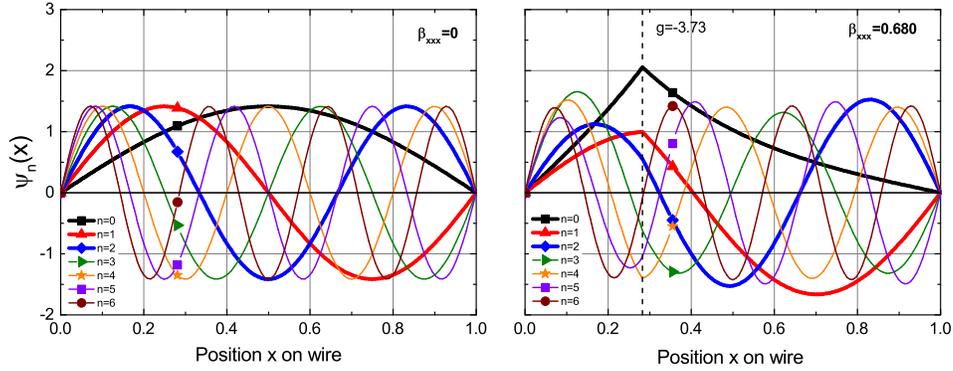}
\caption{Lowest eigenfunctions for a bare wire graph with zero potential (left) and one with a strong, negative $\delta$ potential (right) positioned to maximize $\beta_{xxx}$.}
\label{fig:1deltaWireZeroBetaBestBetaSide2side}
\end{figure}

\begin{figure}\center
\includegraphics[width=5in]{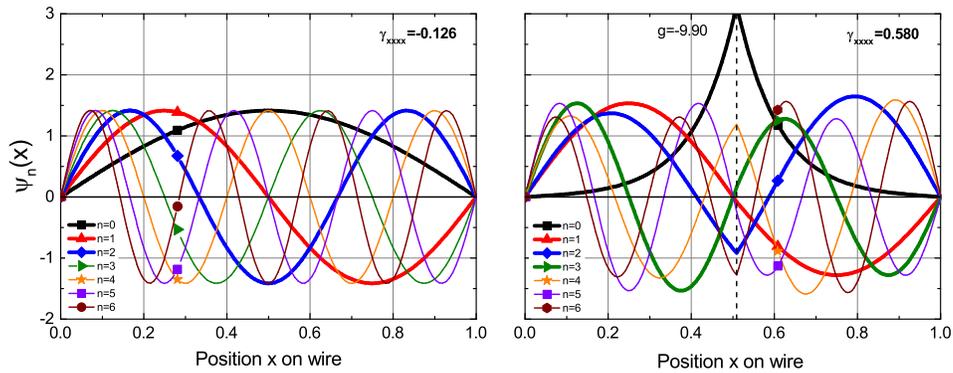}
\caption{Lowest eigenfunctions for a bare wire graph with zero potential (left) and one with a strong, negative $\delta$ potential (right) positioned to maximize $\gamma_{xxxx}$.}\label{fig:1deltaWireLowestGammaBestGammaSide2side}
\end{figure}

The upper right panel in Figure \ref{fig:DeltaWireGraphs} shows a wire to which a finite potential of the form $V(x)=(g/L)\delta (x-a)$ has been placed at location $x=a$.  The nonlinear optics of this graph have been discussed extensively\cite{lytel13.04}.  The $\delta$ function is a source or sink of probability flux, creating a slope discontinuity in the eigenfunctions of the wire.  The right panel in Figure \ref{fig:1deltaWireZeroBetaBestBetaSide2side} shows the localization of the ground state wavefunction to the position of the potential, with little to no change in the localization of the first few excited states.  This delocalization of the ground state from the first few excited states drives the nonlinearities from zero to values near the fundamental limits of the off-resonance hyperpolarizabilities\cite{kuzyk00.01}.  This is even more evident in  for Figure \ref{fig:1deltaWireLowestGammaBestGammaSide2side} for $\gamma_{xxxx}$.

\begin{figure}\center
\includegraphics[width=2.5in]{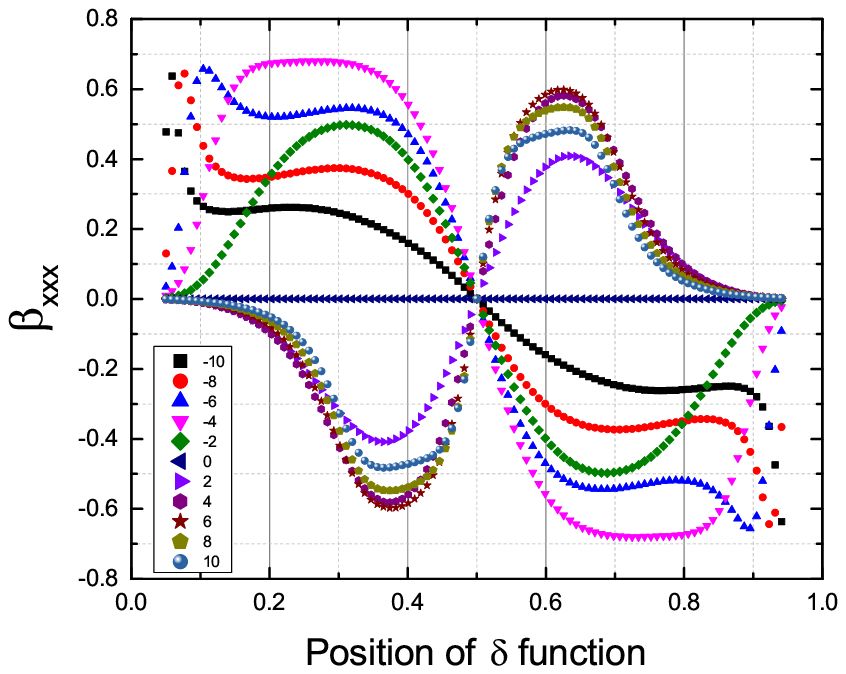}\includegraphics[width=2.5in]{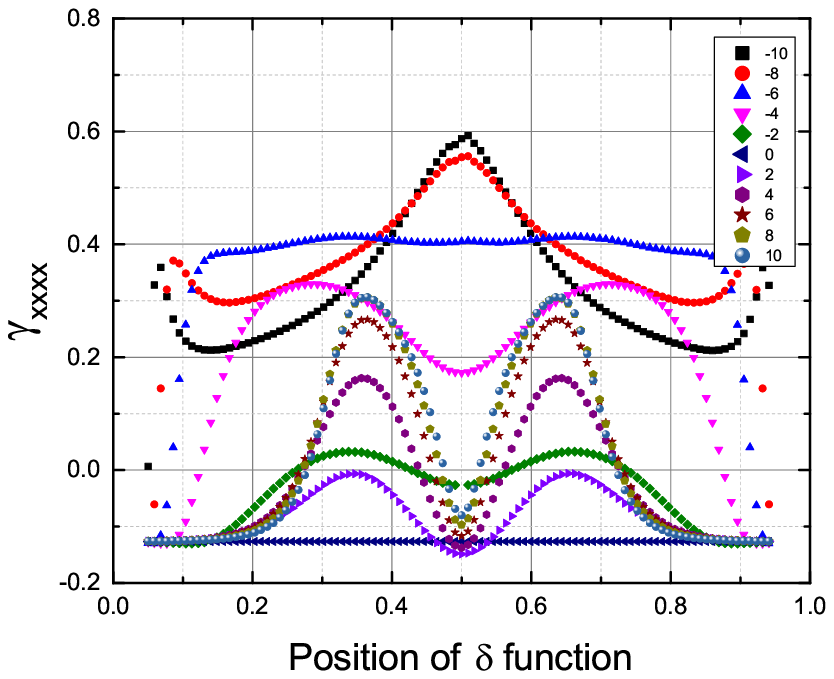}
\caption{Variation of $\beta_{xxx}$ and $\gamma_{xxxx}$ with position along the wire of the $\delta$ function for various strengths.}\label{fig:1delta_wire_hyper_vs_a_at_g}
\end{figure}

Figure \ref{fig:1delta_wire_hyper_vs_a_at_g} shows the hyperpolarizabilities varying with position of the $\delta$ function on the wire at different strengths g.  The spectrum changes when the strength g is smaller than a critical strength $g_c$ determined by the location of the $\delta$ function\cite{blume06.01}.  For $g<g_c<0$, the ground state is negative, and the deep well \emph{pulls} the lowest eigenfunctions to the $\delta$ function, creating a large overlap and generating a large response for $\beta_{xxx}$, even as the $\delta$ function moves toward one end of the wire.  For positive g, $\beta_{xxx}$ gradually goes toward zero as the $\delta$ function moves toward either end.  In both cases, $\beta_{xxx}$ vanishes when the $\delta$ function is at either end, due to the centrosymmetry of the resultant wire.  For $\gamma_{xxxx}$, large values are obtained as the $\delta$ function moves toward the middle of the wire when the well is very deep.

\begin{figure}\center
\includegraphics[width=5in]{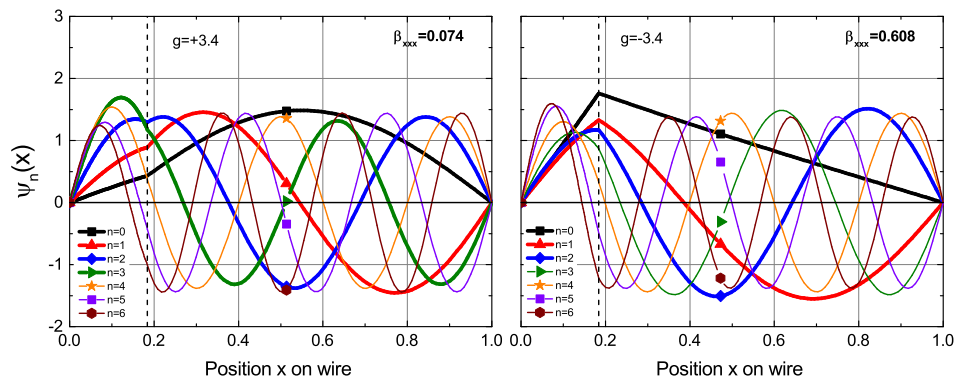}
\caption{First seven eigenfunctions for the $\delta$ graph with positive strength g, near one edge (left), with a small $\beta_{xxx}$, and equal but negative strength g, with a large $\beta_{xxx}$.}\label{fig:1delta_states_poor_max_beta_side2side}
\end{figure}

Figure \ref{fig:1delta_states_poor_max_beta_side2side} shows the eigenfunctions for a graph with a barrier and one with a well of equal potential.  The former looks like a bare wire, with near-zero response, while the latter achieves optimum response near the fundamental limits.  To the extent that a negative $\delta$ function mimics a quantum defect causing disruption of electron flow, the Figure suggests that the presence of a simple defect on a linear structure is sufficient to displace the lowest eigenfunctions to create a large nonlinear optical response.

\begin{figure}\center
\includegraphics[width=3.4in]{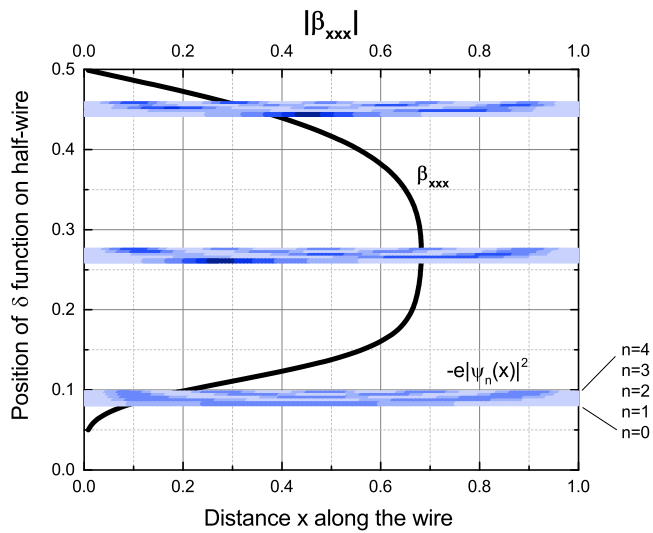}
\caption{Electron charge density for the first five eigenfunctions (with the ground state at the bottom of each band) along the length of the $1\delta$ graph when the $\delta$ function with $g=-3.73$ is placed at three different locations.}\label{fig:electronDensity_1delta_on_wire}
\end{figure}

Figure \ref{fig:electronDensity_1delta_on_wire} shows the relationship between eigenfunction localization and large response.  In the Figure, the electron density across the wire is shown as a shaded set of bars corresponding to the first five eigenfunctions, positioned at three locations of the $\delta$ function on a half-wire.  Highest density has the greatest shading.  Also shown is the value of $\beta_{xxx}$ as a function of the position of the $\delta$ function.  The lowest response occurs when the $\delta$ function (a well, in this case, since $g=-3.73$) is near one end of the wire or at the middle of the wire.  The greatest shift in charge density between the ground and lowest excited states occurs when the $\delta$ function is located a little over a quarter of the way along the wire.  The electron density is delocalized, and $\beta_{xxx}$ is small.

The analysis of the one $\delta$ graphs suggests that quantum structures with defects located at precise points along a wire can produce large enhancements of the nonlinear optical response.  A positive $\delta$ function yields a large response if it is about midway between the center and edge of the structure; a negative $\delta$ function has its greatest effect closer to one end of the structure.  Modulation of the location of the defect by an external \emph{control} electric or magnetic field would cause a huge change in the nonlinear optical refractive index, as is required for an optical modulator or bistable optical switch.

\subsection{Two $\delta$s}\label{2deltaWire}

\begin{figure}\center
\includegraphics[width=5in]{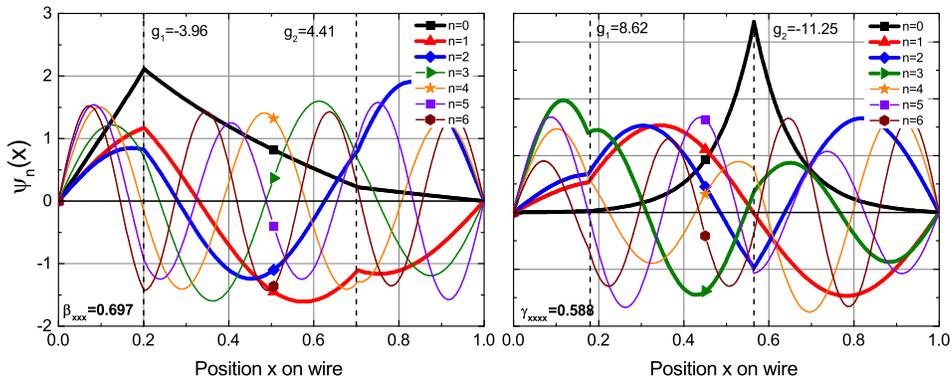}
\caption{First seven eigenfunctions for the two $\delta$ graph with the largest $\beta_{xxx}$ and $\gamma_{xxxx}$.}\label{fig:2deltaStates_best_betaGamma_side_by_side}
\end{figure}

A single $\delta$ function on a wire causes a large change in the nonlinear optical response, generating intrinsic nonlinearities near the apparent fundamental limits.  The lower right panel in Figure \ref{fig:DeltaWireGraphs} shows a wire with two $\delta$ functions on it.  This graph was previously solved\cite{lytel13.04}.  The eigenfunctions for the graphs with the largest $\beta_{xxx}$ and $\gamma_{xxxx}$ are shown in Figure \ref{fig:2deltaStates_best_betaGamma_side_by_side}.

\begin{figure}\center
\includegraphics[width=2.5in]{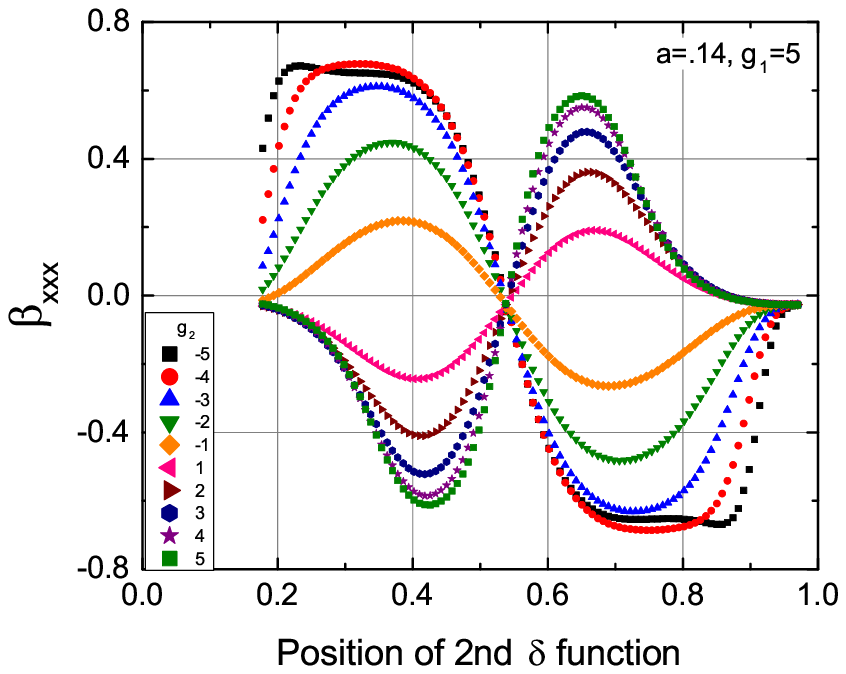}\includegraphics[width=2.5in]{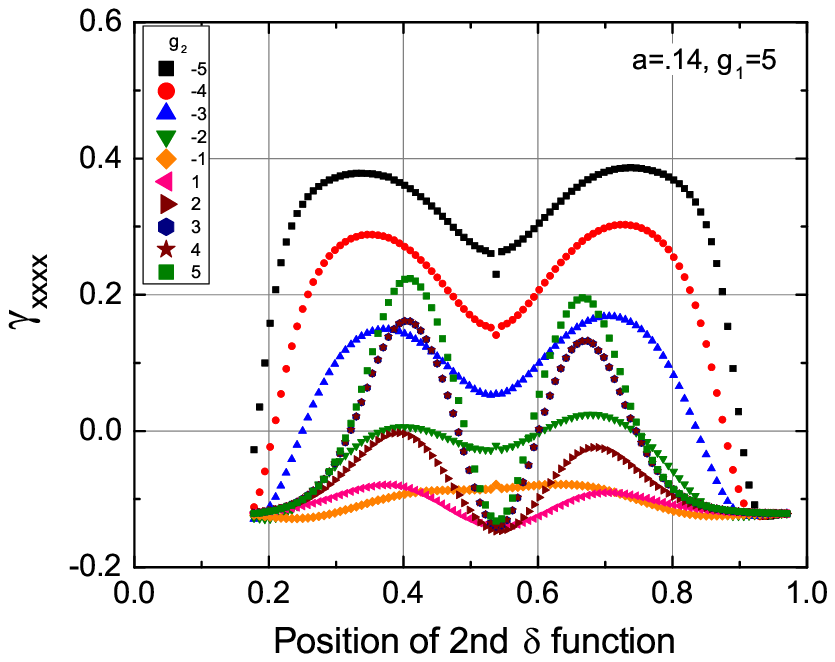}
\caption{Line plots for $\beta_{xxx}$ at $a=0.14, g_1=5$.  This corresponds to a $\delta$ barrier near the left end of the wire.  The various curves show how $\beta_{xxx}$ and $\gamma_{xxxx}$ change as the potential from the second $\delta$ function is moved toward the other end and at various strengths corresponding to both barriers and wells.}\label{fig:beta_gamma_vs_b_at_g2_fixed_a_fixed_g1}
\end{figure}

\begin{figure}\center
\includegraphics[width=2.5in]{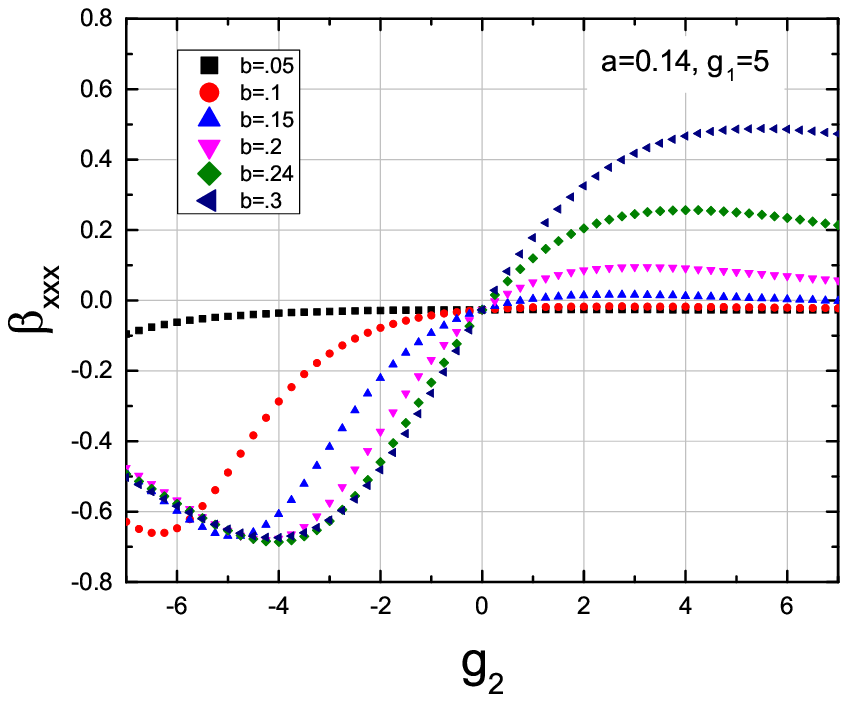}\includegraphics[width=2.5in]{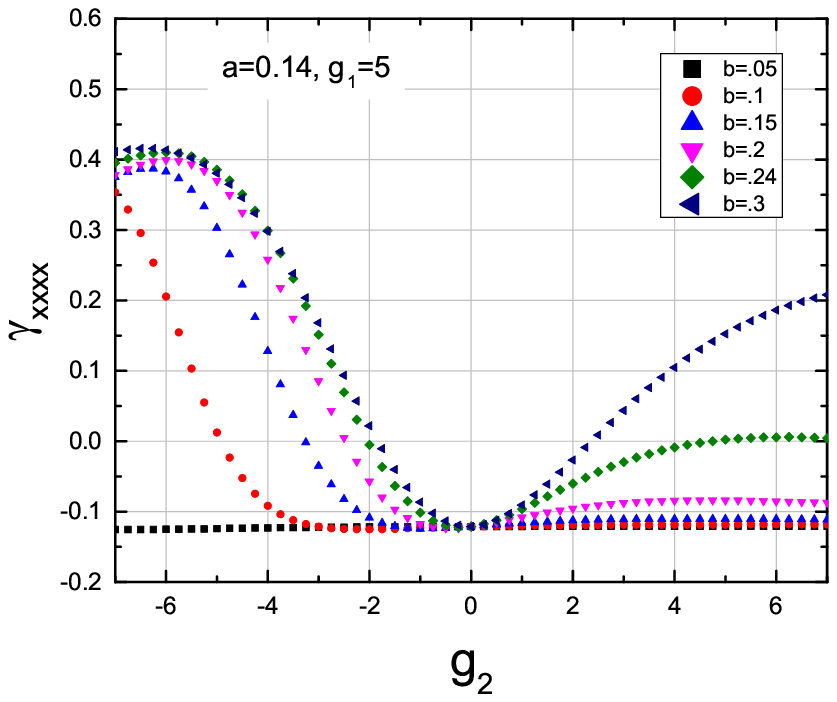}
\caption{Line plots for $\beta_{xxx}$ for $a=0.14, g_1=5$ as the second $\delta$ function is fixed at several distances from the right edge, with varying strengths.}\label{fig:mol2_beta_gamma_vs_g2_at_b}
\end{figure}

Figure \ref{fig:beta_gamma_vs_b_at_g2_fixed_a_fixed_g1} displays the dependence of the hyperpolarizabilities when one $\delta$ function is fixed and the other scans across the wire.  Figure \ref{fig:mol2_beta_gamma_vs_g2_at_b} displays the dependence of the hyperpolarizabilities when one $\delta$ function is fixed near the left edge ($a=0.14)$ with fixed strength $g_{1}=5$ and the other is positioned at a fixed distance b from the right edge, but for several values of $g_{2}$.  These are essentially vertical slices in the prior Figure and show that the combination of a barrier near one end and a well at the other results in a large response for either hyperpolarizability, a toy donor-acceptor model for a quantum wire molecule.

\begin{figure}\center
\includegraphics[width=2.5in]{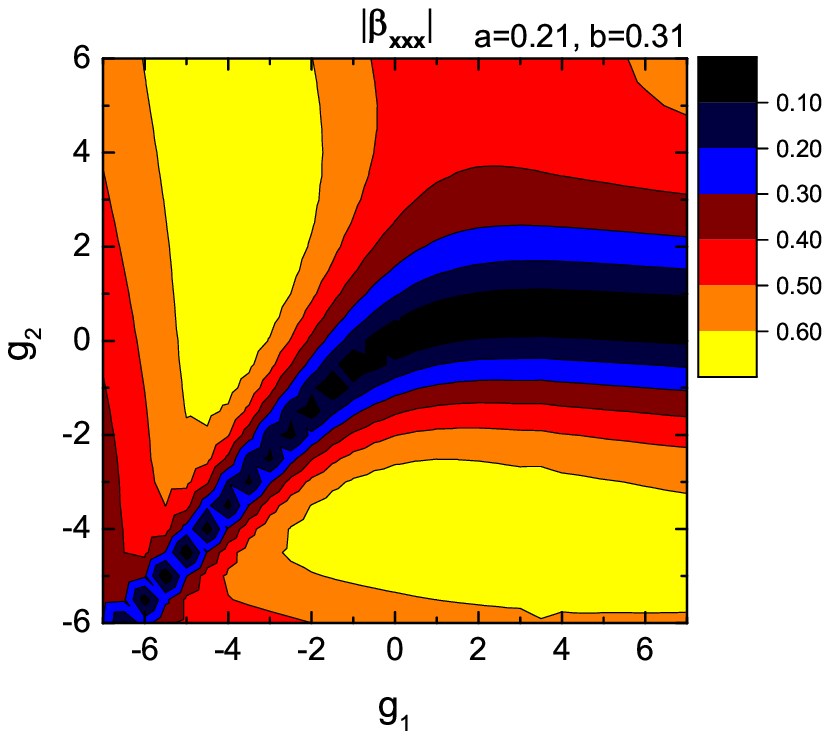}\includegraphics[width=2.5in]{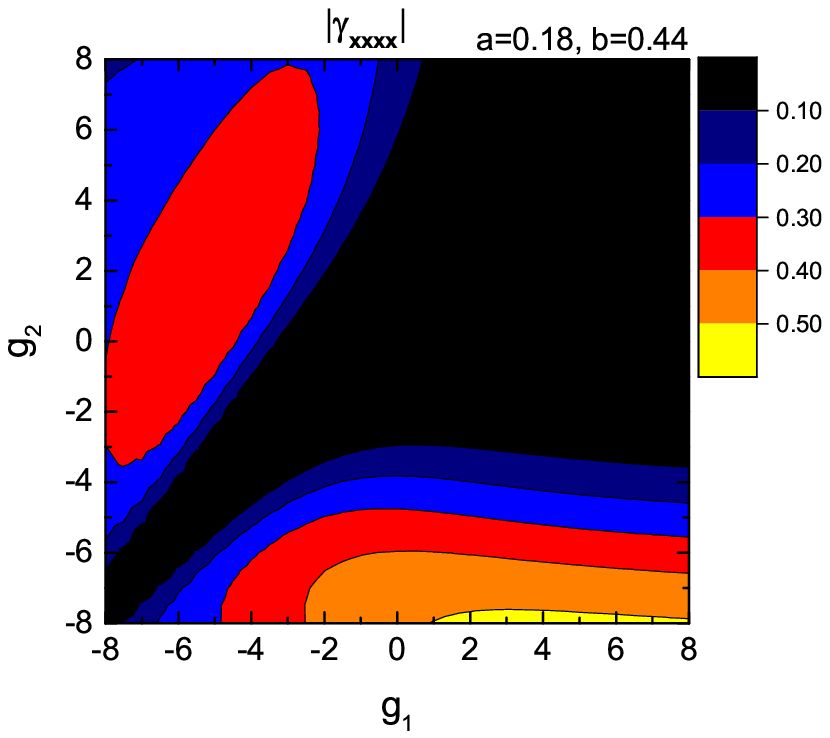}
\caption{Contour plot for the two $\delta$ graph of $\beta_{xxx}$ and $\gamma_{xxxx}$ for fixed $\delta$ locations but variable strengths.}\label{fig:2delta_BestbetaGamma_vs_g1g2_fixed_a_b_contour}
\end{figure}

Figure \ref{fig:2delta_BestbetaGamma_vs_g1g2_fixed_a_b_contour} show the dependence of the hyperpolarizabilities on the strengths of the two $\delta$ functions.

The results of this work indicate that an additional defect has little effect on the nonlinear optical response, once a single defect is in place at the proper location.

\subsection{Three $\delta$s}\label{3deltaWire}

\begin{figure}\center
\includegraphics[width=5in]{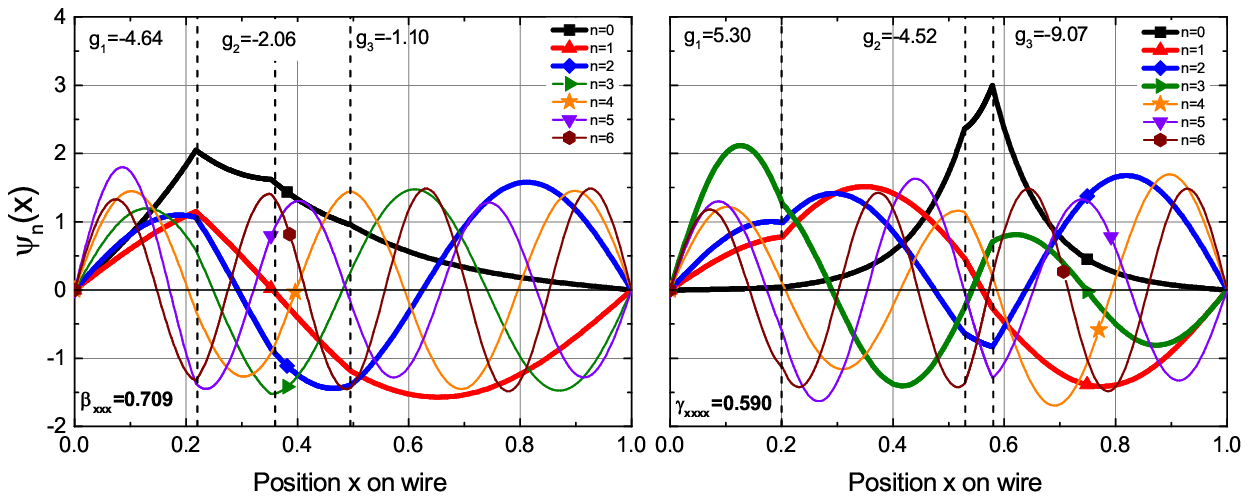}
\caption{First seven eigenfunctions for the three $\delta$ graph with the largest $\beta_{xxx}$ and $\gamma_{xxxx}$.}\label{fig:3deltaStates_best_betaGamma_side_by_side}
\end{figure}

The lower right panel of Figure \ref{fig:DeltaWireGraphs} shows three $\delta$ functions on a wire.  This graph was previously studied in detail\cite{lytel13.04}.  The eigenfunctions for the graphs with the largest $\beta_{xxx}$ and $\gamma_{xxxx}$ are shown in Fig \ref{fig:3deltaStates_best_betaGamma_side_by_side}.

Prior analysis showed that the hyperpolarizabilities approached their maximum values with one $\delta$ function with the proper placement and strength\cite{lytel13.04} .  The maximum values were just shy of the theoretical maxima implied by the theory of fundamental limits\cite{kuzyk00.01}.  The addition of a second $\delta$ function produced a very slight increase in both hyperpolarizabilities, and we see that the three $\delta$ wires have hyperpolarizabilities that are essentially at the fundamental limits. We conclude that it is not necessary to interrupt the flux on a donor-acceptor chain with more than one $\delta$ function or prong.  This is in agreement with prior analysis showing that only a few parameters are required to achieve near-maximum nonlinearities, despite that this poorly defines the potential\cite{atherton12.01}.

Again, the presence of an additional defect does little to enhance the nonlinearities, but it also does not negatively affect them.  This would suggest that a wire with many randomly placed $\delta$ defects, would likely have at least one in a position generating a large response, and that the presence of the rest of the random defects would have minimal positive \emph{or} negative impact.

\section{Quantum structures with side groups}\label{sec:prong}

\begin{figure}\center
\includegraphics[width=2.5in]{Wire_bare.eps}\includegraphics[width=2.5in]{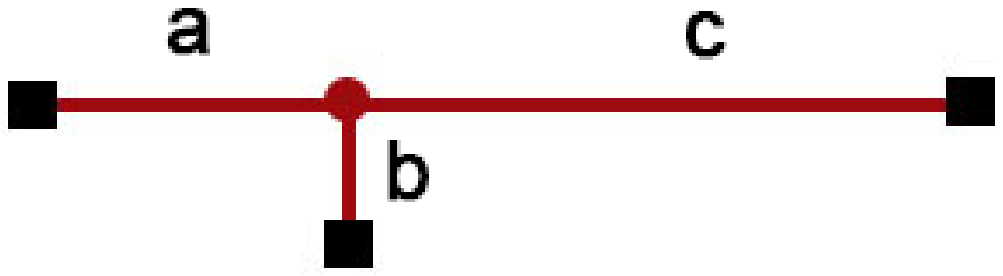}\\
\includegraphics[width=2.5in]{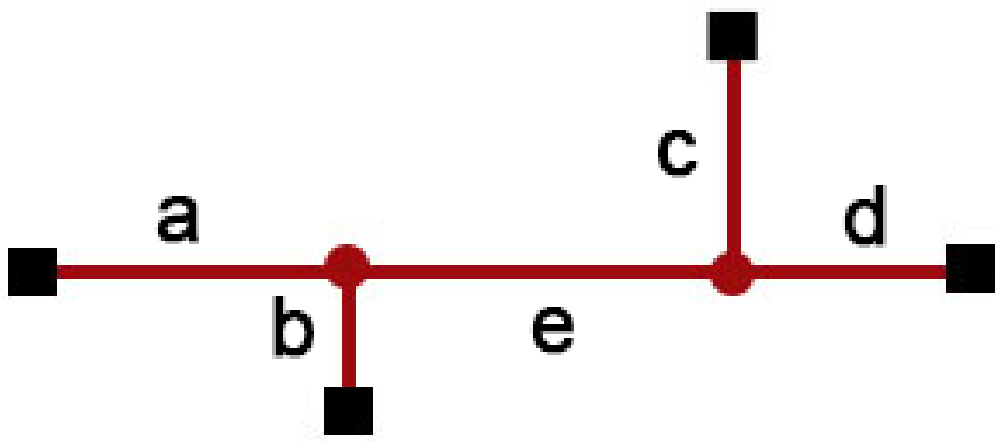}\includegraphics[width=2.5in]{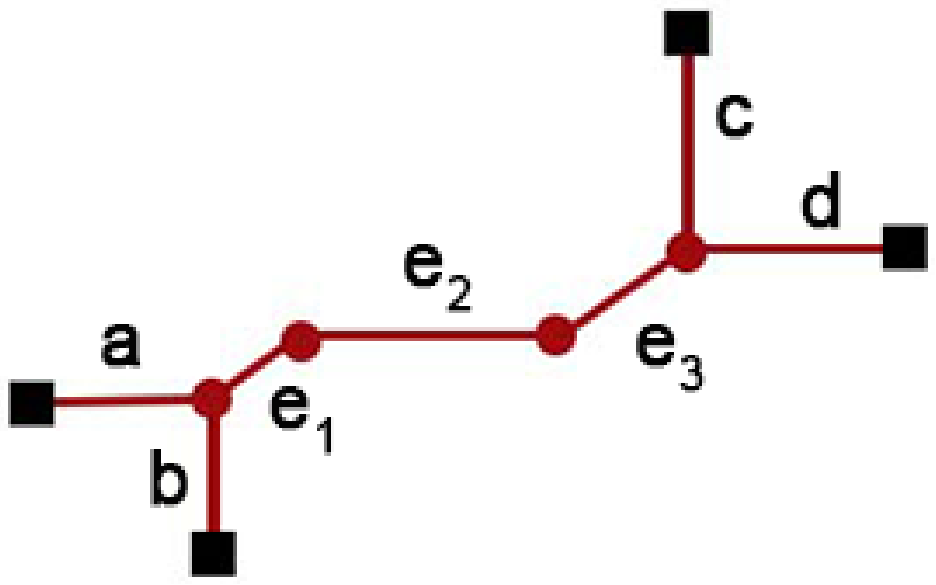}\\
\caption{The bare wire graph and a variant with one prong, one with two prongs, and another with two prongs but with a geometry suggestive of a donor-acceptor molecule. The probability flux at the position of the prongs is conserved but the prong carries away some of the probability flux, creating an apparent discontinuity across the straight part of the wire at the prong location.}
\label{fig:ProngWireGraphs}
\end{figure}

Figure \ref{fig:ProngWireGraphs} shows a bare wire graph and three variants, this time with prongs attached.  The addition of a prong at a junction to a bare wire creates a 3-star graph, previously studied in detail for its quantum mechanical properties\cite{pasto09.01} and solved for its nonlinear optical properties\cite{lytel13.02}.  One of the variants has the geometry of the series of donor-acceptor structures\cite{may07.01}.  The presence of a prong creates a pathway for charge flow away from the main chain.  This changes the spectrum to a quasi-random set of energies lying within fixed root boundaries equal to the energies of the bare wire. It also changes the shape of the eigenfunctions, affecting the change in dipole moment and the off-diagonal transition moments.  The proper choice of a prong length and position can lead to a large enhancement in the hyperpolarizabilities.

\subsection{Wire with one prong}\label{1prongWire}

\begin{figure}\center
\includegraphics[width=2.5in]{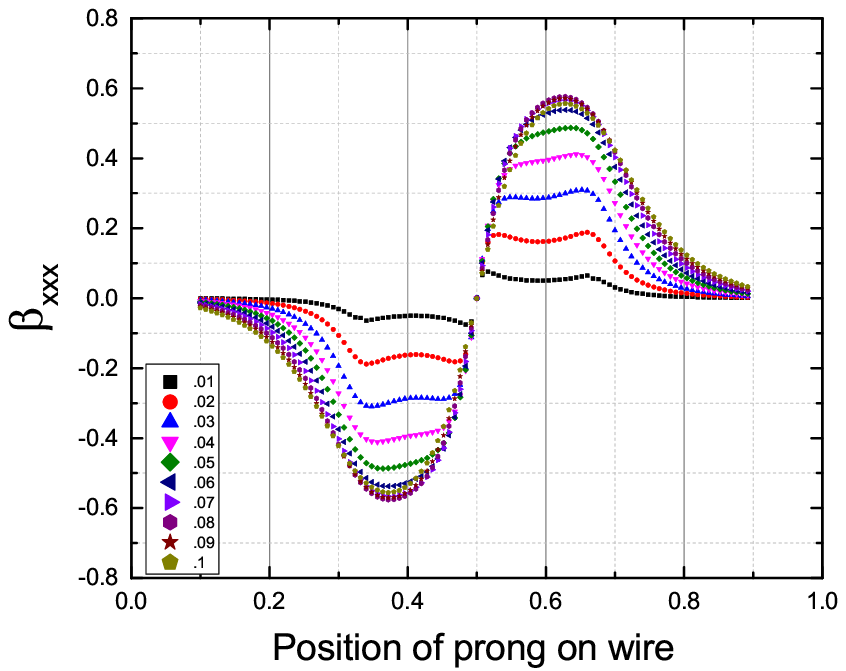}\includegraphics[width=2.5in]{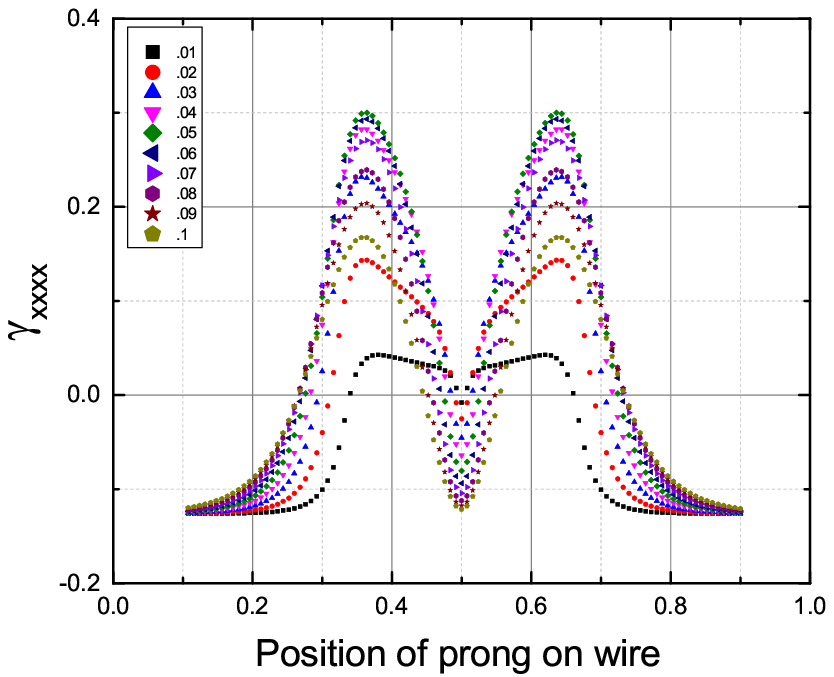}
\caption{Variation of $\beta_{xxx}$ and $\gamma_{xxxx}$ with position of the prong for various prong lengths.}\label{fig:1prong_wire_hyper_vs_a_at_b}
\end{figure}

The upper right panel of Figure \ref{fig:ProngWireGraphs} shows a bare wire graph to which a single prong has been attached instead of a $\delta$ function.  This graph has a star topology.  Figure \ref{fig:1prong_wire_hyper_vs_a_at_b} show the variation of the hyperpolarizabilities with position of the prong on the wire at a number of prong lengths.  The behavior resembles that of the $\delta$ graph when $g>0$.  There is no analog to the negative g $\delta$ graph for the prong graph because the energy spectrum of star graphs is always positive.

\begin{figure}\center
\includegraphics[width=5in]{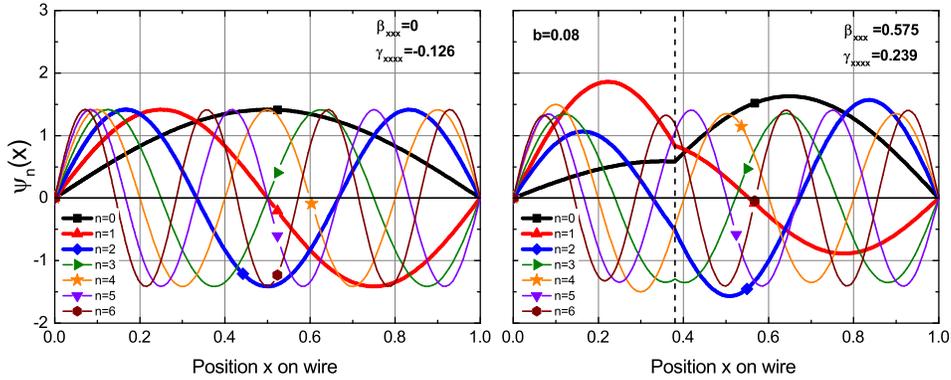}
\caption{Lowest eigenfunctions for a bare wire graph (left) and one with a prong of length $b=0.08$ positioned to maximize $\beta_{xxx}$  This configuration has a large $\gamma_{xxxx}$ as well.  If the prong length is decreased to $b=0.05$, then $\gamma_{xxxx}\rightarrow 0.300$, its maximum for the one prong graph, while $\beta_{xxx}$ drops slightly. }
\label{fig:1prong_states_zeroBetaBestBetaSide2side}
\end{figure}

Figures \ref{fig:1prong_states_zeroBetaBestBetaSide2side} compares the first seven eigenfunctions of the bare wire graph (left) with those of the one prong graph with the largest $\beta_{xxx}$.  The enhancement induced by the existence of a short prong arises from the change induced in the distribution of electron flux in the ground state, compared to the relatively unchanged shape of the first few excited states.

\begin{figure}\center
\includegraphics[width=2.5in]{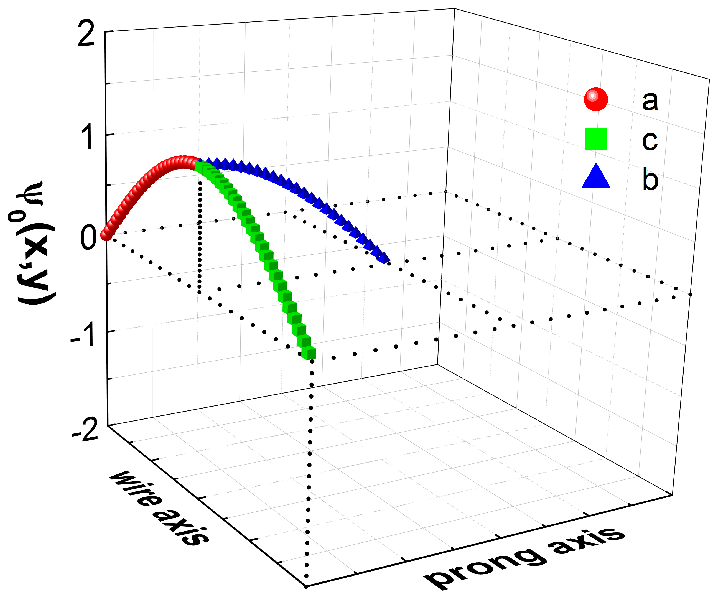}\includegraphics[width=2.5in]{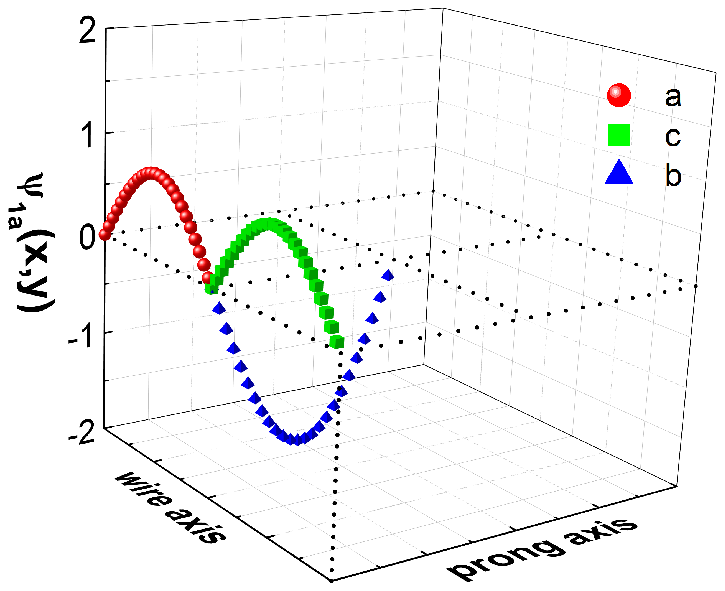}\\
\includegraphics[width=2.5in]{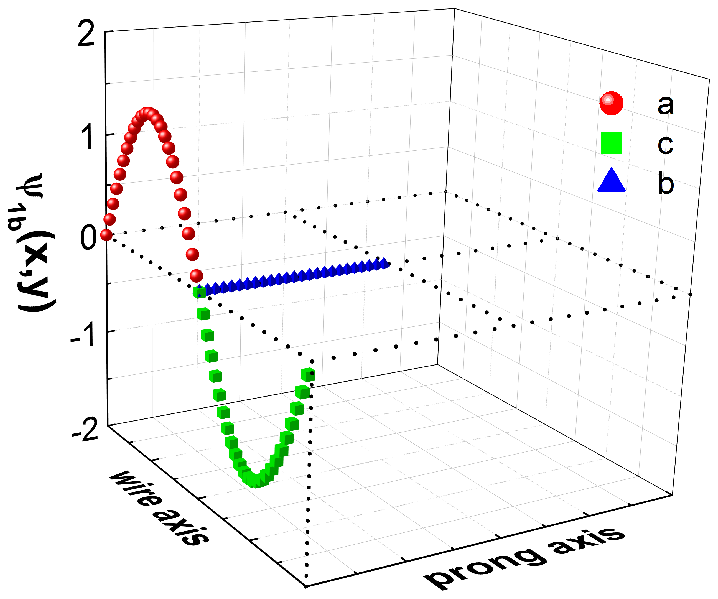}\\
\caption{Three-dimensional view of the edge states for the wire with one prong.  In this graph, the two edges a and c are equal, and the prong is labeled b.  The first excited state is two-fold degenerate, as shown.  For this graph, $\beta_{xxx}$ is nonzero but small.  The scale on the edges is arbitrary, as the intrinsic nonlinearities are scale-independent, but it is convenient to choose $a+c=1$ along the wire axis.  The vertical axis is labeled with the value of the amplitude of the normalized eigenfunctions.}\label{fig:1prong3Ddegenerate}
\end{figure}

Figure \ref{fig:1prong3Ddegenerate} is a 3D plot of the edge vectors for the 1 prong graph for the first three eigenfunctions.  The two edges along the x-axis are of equal length, producing a degeneracy in the first excited state, as shown in the Figure.  One degenerate state has a nonzero amplitude on all edges, while the other has exactly zero flux on the prong.  The nonlinearities for this graph are well below the maximum values for this topology.

\begin{figure}\center
\includegraphics[width=2.5in]{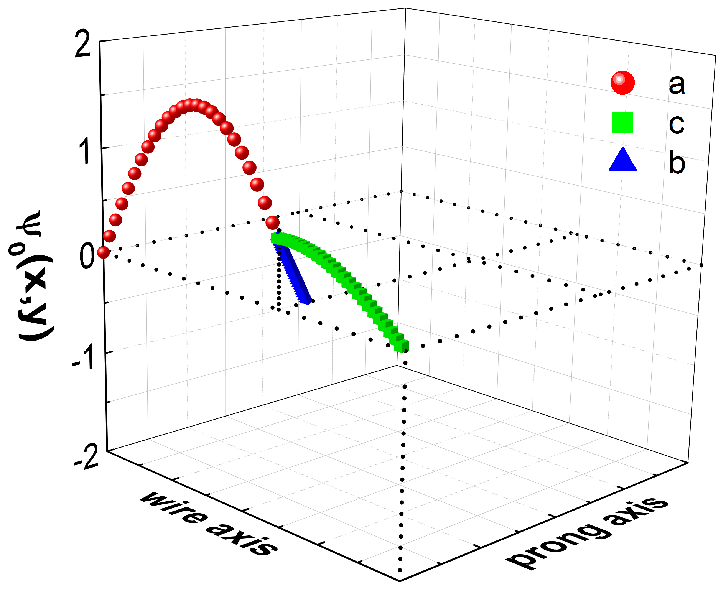}\includegraphics[width=2.5in]{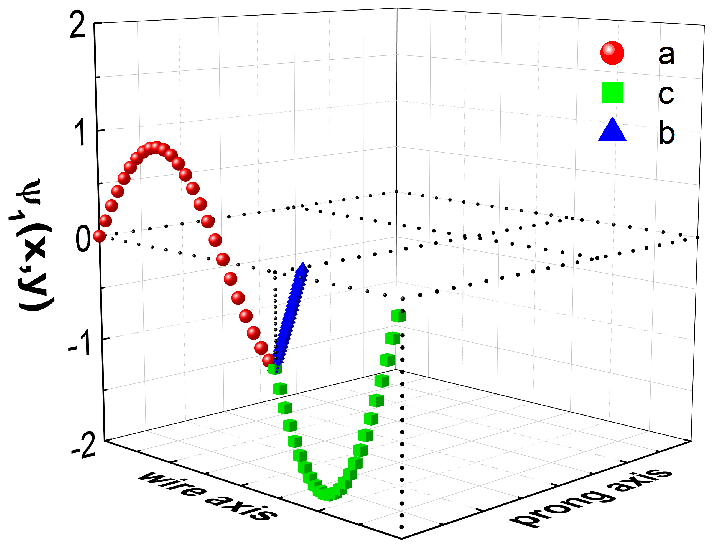}\\
\includegraphics[width=2.5in]{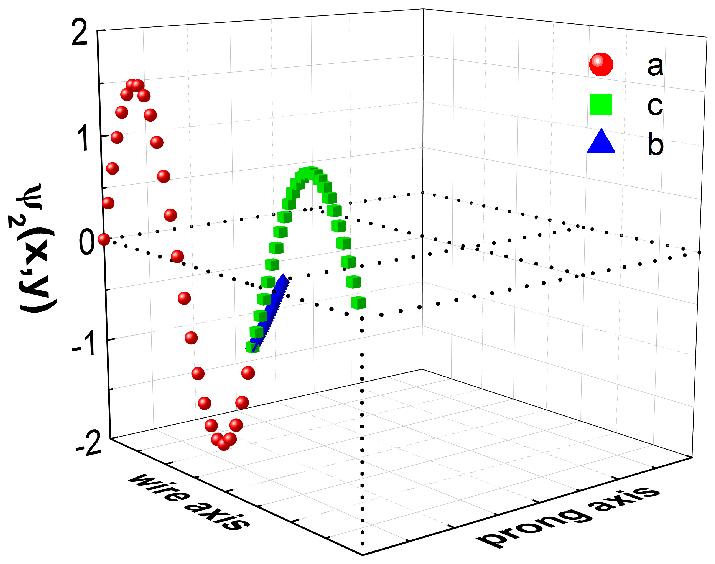}\\
\caption{Three-dimensional view of the edge states for the wire with one prong.  In this graph, the two edges a and c are unequal, and all eigenfunctions are singlets. The prong interrupts the flux along the wire and generates the maximum $\beta_{xxx}$ for this topology.}\label{fig:1prong3DmaxBeta}
\end{figure}

Figure \ref{fig:1prong3DmaxBeta} is a 3D plot of the edge vectors for the 1 prong graph for the first three eigenfunctions when the two edges along the x-axis are irrationally-related, producing a nondegenerate spectrum.  The $\beta_{xxx}$ for this graph is at the maximum values for this topology.  The effect of the prong is to cause a large slope discontinuity along the edge, localizing the lowest order states and leading to a large response.

How similar are prong graphs and $\delta$ graphs with $g>0$?  The secular equation for the one prong graph may be written as
\begin{equation}\label{secStar}
\cot{ka}+\cot{kc}+\cot{kb}=0.
\end{equation}
The secular equation for the one $\delta$ graph may be written as
\begin{equation}\label{secDelta}
\cot{ka}+\cot{kc}+\frac{2g}{kL}=0.
\end{equation}
The equality of these two spectral equations would require an energy-dependent $g(k)$ given by $g(k)=\frac{kL}{2}\cot{kb},$ but when k is near zero, $g(k)\rightarrow \frac{L}{2b}$ and becomes energy-independent.  For $b=0.08$, we find near-equality in the states and spectra when $g\sim 6.$  Thus prong graphs and $\delta$ graphs are essentially equivalent in this case.

To first order, the position for a prong on a wire that leads to maximum $\beta_{xxx}$ is evidently determined by maximizing the product $|x_{01}| ^{2}\bar{x}_{11}$.  Moving the prong from the edge toward the center pushes the ground and first excited states to opposite sides of the prong, which increases the change in dipole moment but decreases the ground to first excited state transition moment.  The ideal position is somewhere in-between the edge and the center.  The conclusion we can draw is that shaping the eigenfunctions in this manner will produce a large, intrinsic first hyperpolarizability.  Similar reasoning holds for $\gamma_{xxxx}$ when $g>0.$.

\begin{figure}\center
\includegraphics[width=2.5in]{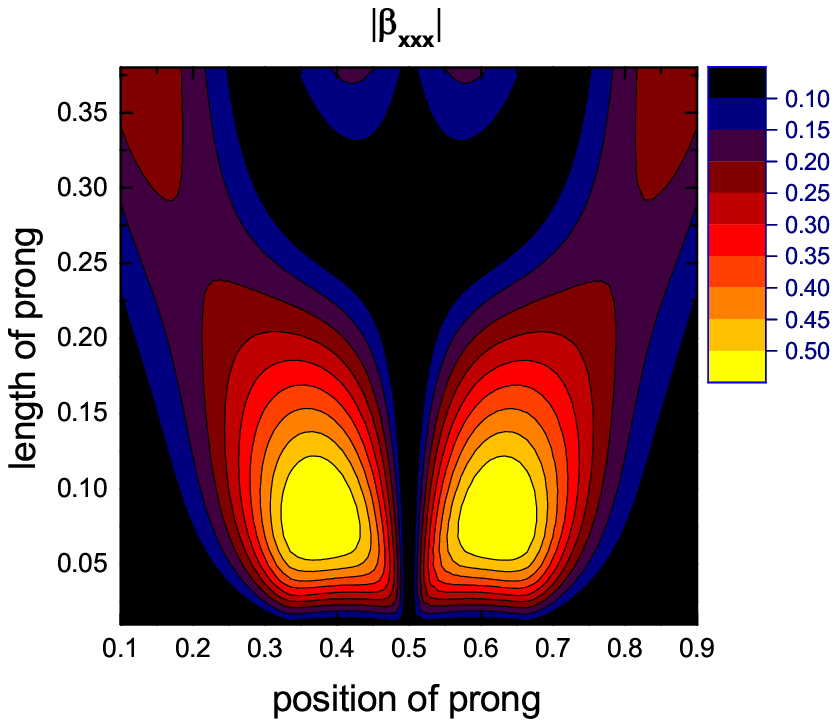}\includegraphics[width=2.6in]{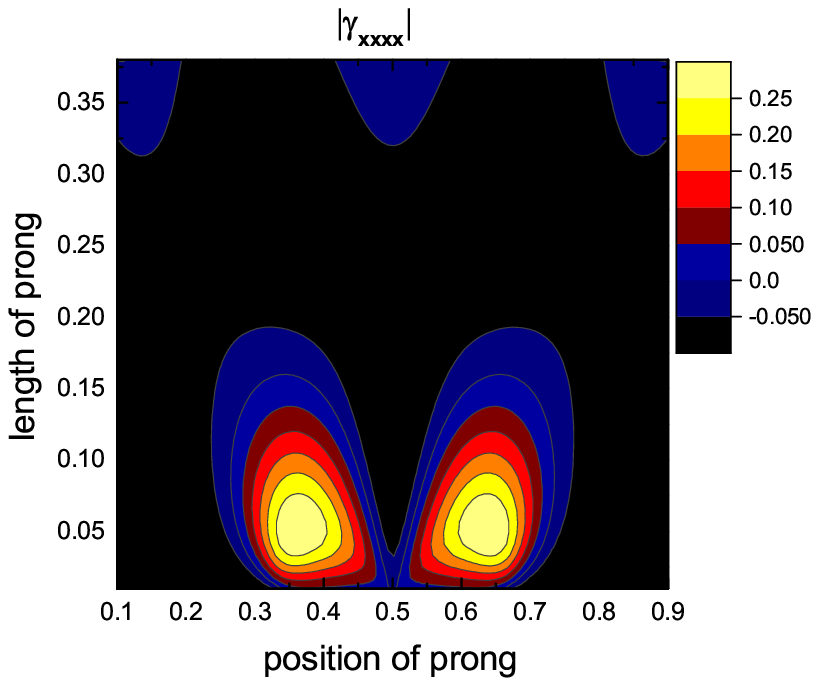}\\
\caption{Contour plot of $|\beta_{xxx}|$ over the range of the position and length of the prong.}\label{fig:1prong_betaGamma_map_over_a_b_contour}
\end{figure}

It appears that the genesis of large nonlinearities is a distortion of the eigenfunctions along the main chain by the presence of the disturbance at the junction.  The disturbance induces as phase shift in the part of the eigenfunction that lies along the main chain, producing a change in dipole moments sufficient to generate a larger $\beta$ or $\gamma$.  To get a feel for how the one electron graph behaves as the prong length is varied while its position is fixed, see the contour graph in Figure \ref{fig:1prong_betaGamma_map_over_a_b_contour}.
It is evident that $\beta$ is nearly zero for near-zero length prongs, regardless of where the prong is positioned.  For a fixed prong position, say $a=0.37$, increasing the prong length from zero causes a jump in $\beta$ to its maximum for this topology, and continuing to increase the prong length beyond a value of about $0.08$ causes $\beta$ to drop again to a low value.  For a fixed prong length, its ideal position is about a third of the way along the wire, while its least effective position is at either end, or at the center where the structure becomes centrosymmetric and $\beta\rightarrow 0.$

Moreover, the length of the prong attached to the main chain cannot be near-zero, nor can it get too large.  To quantify this, abandon the stick model for a moment and consider the transition moments on the main wire of the chain and those on the side group.  First, since the side group is effectively normal to the main chain, it doesn't contribute to $\beta_{xxx}$ so we don't want much of the transition moment to arise from this group.  This means making it short so that an integral of the eigenfunction over that length is small compared to those on the main chain.  That is precisely the reason we do not want a long side chain.  But if the side chain is too short, the eigenfunction on the side chain develops a very large amplitude over a small distance, creating a steep slope at its junction with the main chain and causing the ground state on the main chain to completely localize to one side of the attachment point and the first few excited states to localize to the other side.  This has the effect of driving the transition moments $x_{01}$ and $x_{02}$ to near-zero values, thus driving $\beta$ and $\gamma$ to zero.

\begin{figure}\center
\includegraphics[width=4in]{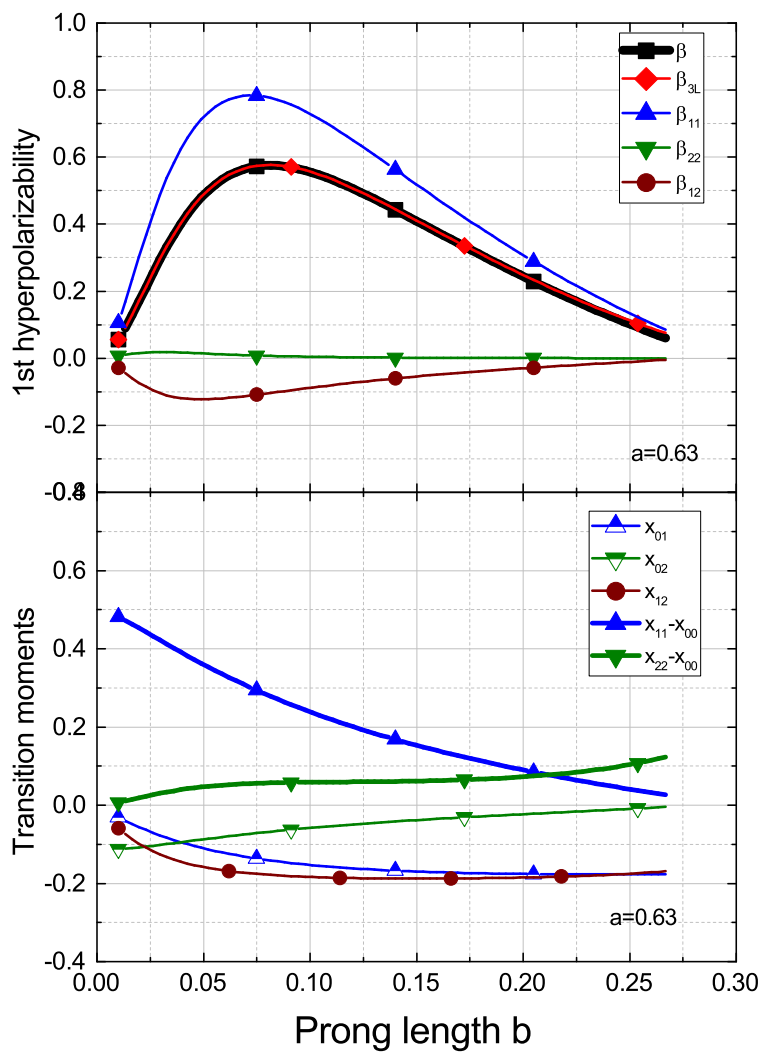}
\caption{Contributions to $\beta_{xxx}$ (top) and the transition moments from which they arise (bottom) as the prong length is varied from small to large.}\label{fig:1prong_beta_and_moments}
\end{figure}

Contributions to $\beta$ (top) and the transition moments from which they arise (bottom) as the prong length is varied from small to large are shown in Figure \ref{fig:1prong_beta_and_moments}. Note that it is $\beta_{11}$ that contributes the most to $\beta_{xxx}.$  So we are most interested in how the contributing moments $\bar{x}_{11}$ and $x_{01}$ change with prong length b.

\begin{figure}\center
\includegraphics[width=5in]{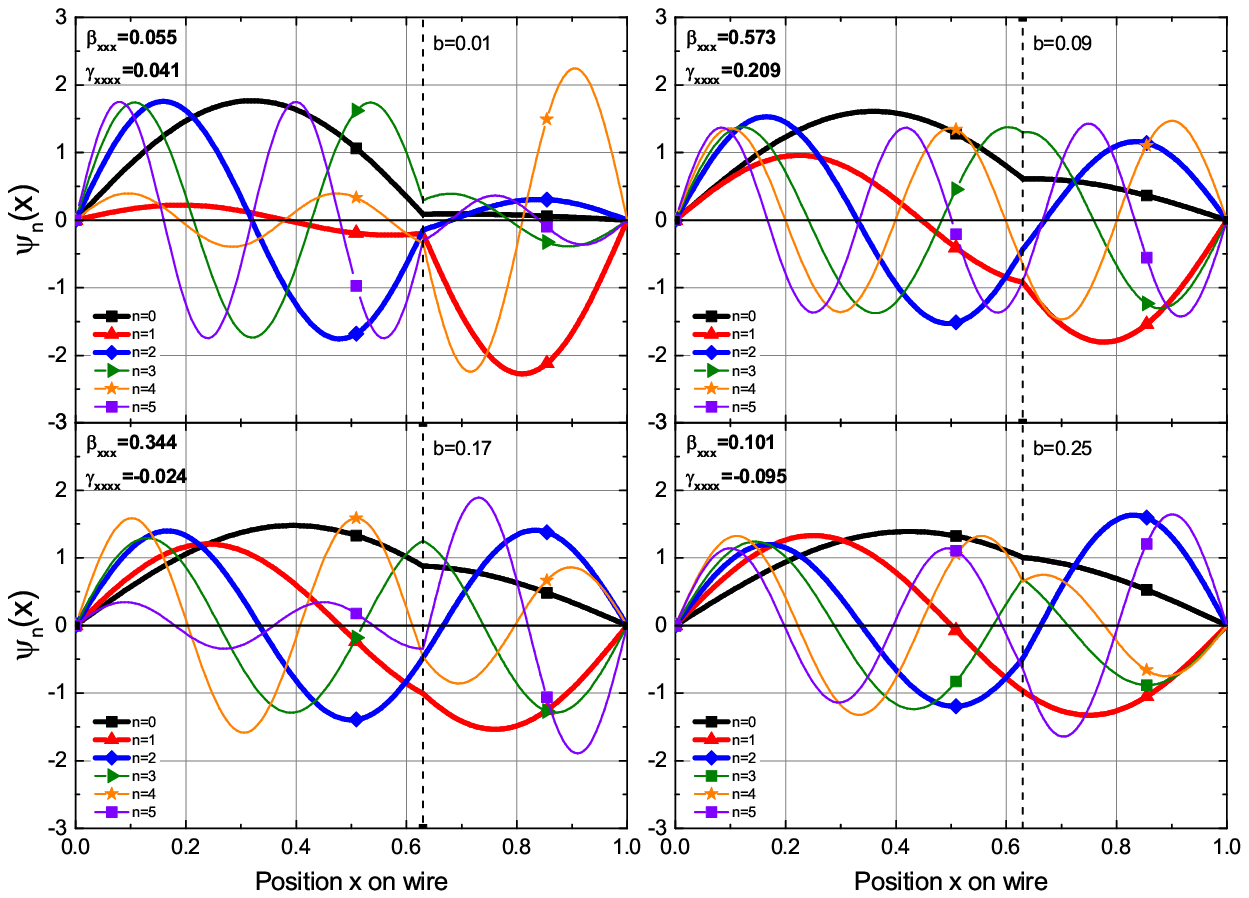}
\caption{Edge states along the main chain for short prongs.}\label{fig:1prong_states_quadPlot}
\end{figure}

Figure (\ref{fig:1prong_states_quadPlot}) shows exactly how the dependence in Figure (\ref{fig:1prong_beta_and_moments}) arises.  In the top left panel, the prong is very short and creates a large change in dipole moment, but the prong is so short that the ground and first excited state are localized to opposite sides of the prong, reducing their overlap transition moment $x_{01}$ to near zero and effectively zeroing $\beta_{xxx}$.  As the prong length is increased (upper right panel), the change in dipole moment is reduced somewhat, but the overlap jumps, and the value of $\beta_{xxx}$ jumps to near its maximum for this topology.  Continuing to increase the prong length keeps the overlap moments large but reduces the change in dipole moment to near zero, thus reducing $\beta_{xxx}$ again to a small value.  The existence of an optimum prong length is highly suggestive of a design rule for side chain lengths on a donor-acceptor chain.  This will be discussed shortly.  The figure also shows how the three level model is nearly exact for $\beta_{xxx}$ for this graph.

\subsection{Two prongs}\label{2prongWire}

The lower left panel of Figure \ref{fig:ProngWireGraphs} shows a wire with two prongs.  This graph is a \emph{back-to-back} star graph, with two star vertices, but with the prongs perpendicular to the main wire\cite{lytel13.04}.

\begin{figure}\center
\includegraphics[width=5in]{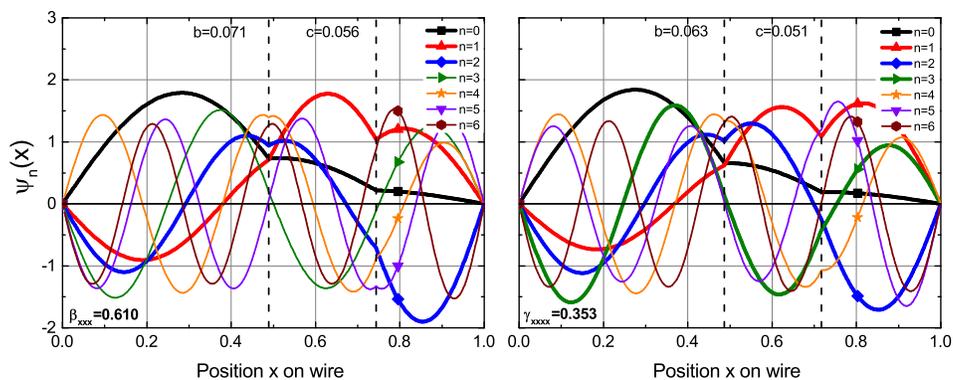}
\caption{First seven eigenfunctions for the two prong graph with the largest $\beta_{xxx}$ and $\gamma_{xxxx}$.}\label{fig:2prongStates_best_betaGamma_side_by_side}
\end{figure}

\begin{figure}\center
\includegraphics[width=2.5in]{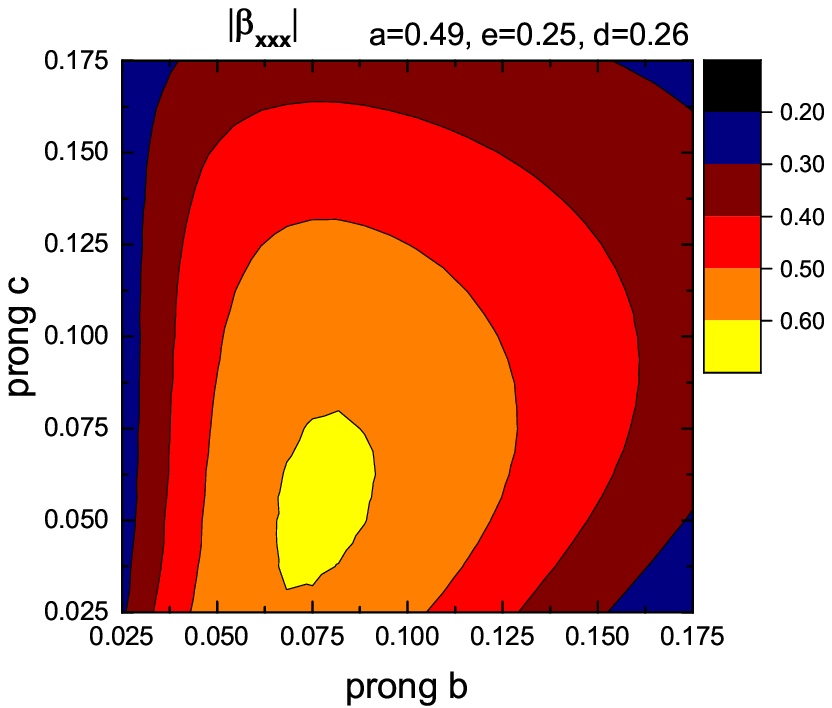}\includegraphics[width=2.5in]{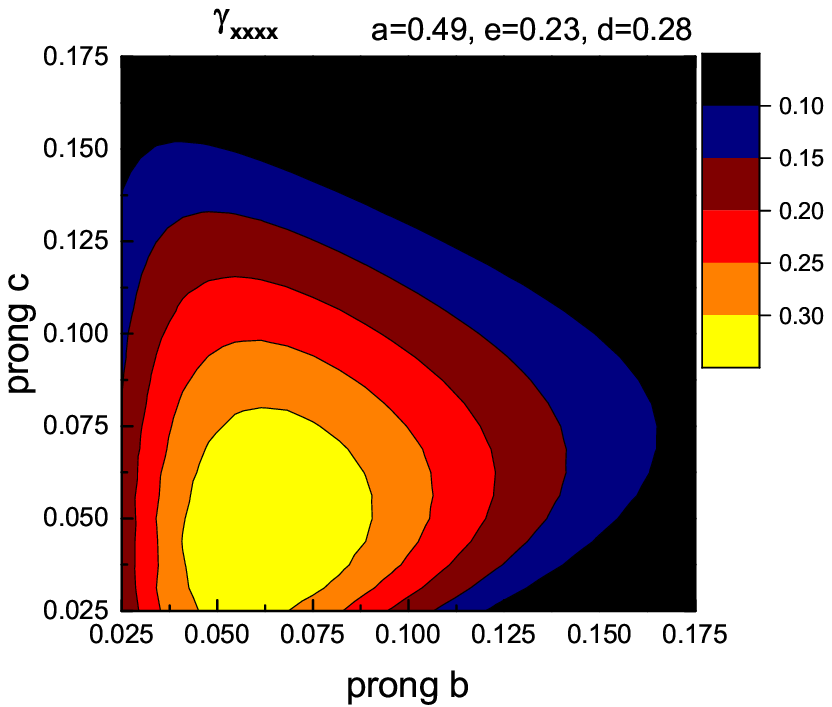}
\caption{Contour plot for the two prong graph of $\beta_{xxx}$ (left) and $\gamma_{xxxx}$ (right) for fixed prong locations but variable lengths.}\label{fig:2prong_beta_gamma_vs_b_c_at_fixed_a_e_d_contour}
\end{figure}

\begin{figure}\center
\includegraphics[width=2.5in]{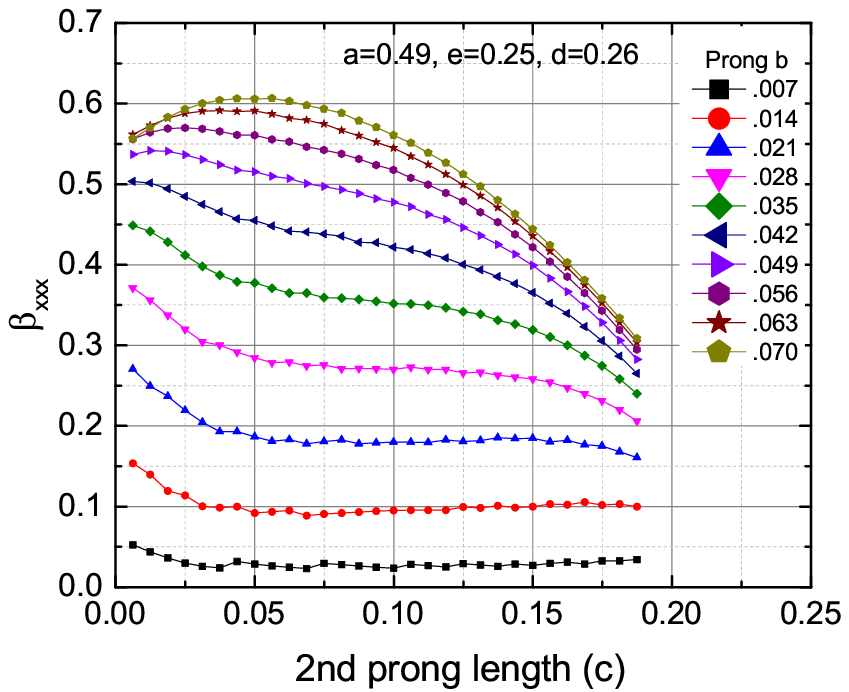}\includegraphics[width=2.5in]{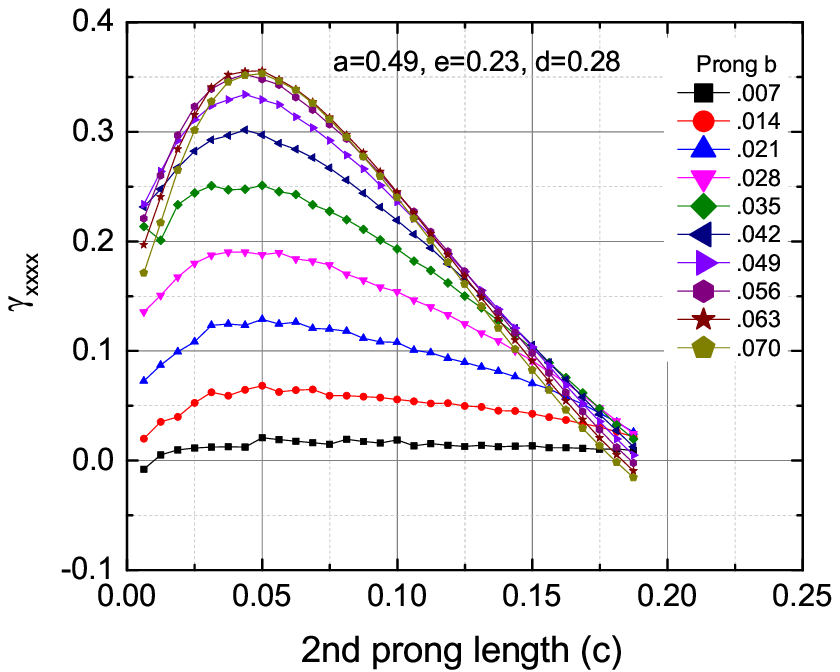}
\caption{Variation of $\beta_{xxx}$ and $\gamma_{xxxx}$ with the length of the second prong c (located at $x=a+e$) for various lengths of the first prong b (located at $x=a$), over a unit length wire.}\label{fig:2prong_beta_vs_b_c_at_fixed_a_e_d_autoparsed}
\end{figure}

Figure \ref{fig:2prongStates_best_betaGamma_side_by_side} shows the eigenfunctions for the two prong graphs with the largest value of the hyperpolarizabilities.
Figure \ref{fig:2prong_beta_gamma_vs_b_c_at_fixed_a_e_d_contour} displays the variations of $\beta_{xxx}$ and $\gamma_{xxxx}$ as each of the two prongs varies in size. One prong has been located near the center of the wire, while the other is midway between the center and the right edge. As is obvious from the eigenfunctions, short prongs are ideal for introducing a flux interruption along the wire, separating the ground state flux from those in the first few excited states.
Figure \ref{fig:2prong_beta_vs_b_c_at_fixed_a_e_d_autoparsed} presents a parametric view of the variation with prong lengths.

\subsection{Two prong wire with bent connector}\label{Biaggio}

\begin{figure}\center
\includegraphics[width=2.5in]{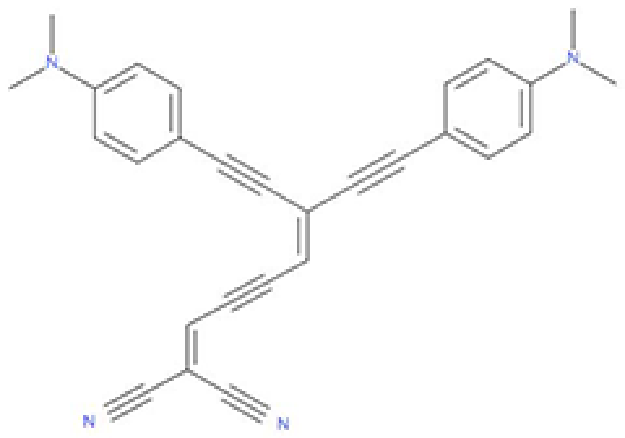}\includegraphics[width=2.5in]{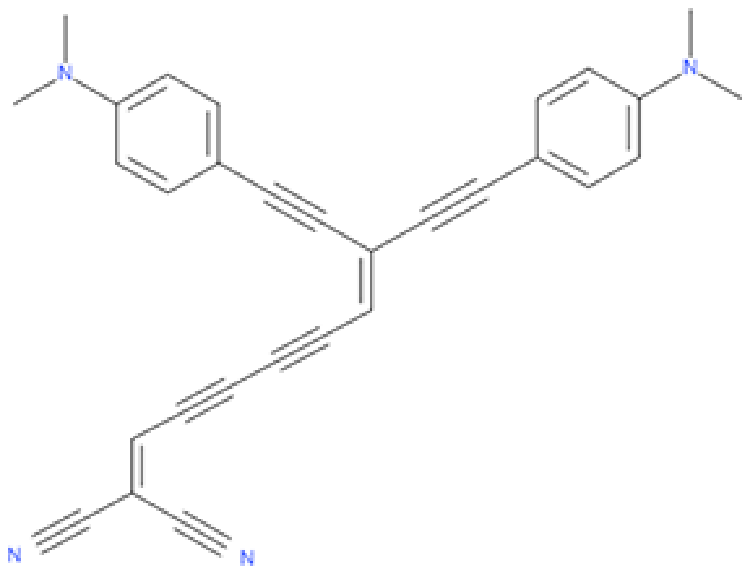}\\
\caption{Molecular form of structures under study.}\label{fig:Biaggio_molecules}
\end{figure}

The lower right panel of Figure \ref{fig:ProngWireGraphs} shows a two prong graph but with a bent wire connecting the two star nodes and with the geometry of the molecules shown in Figure \ref{fig:Biaggio_molecules}.  These represent a subset of small cyanoethynylethene molecules in which donor-acceptor substitution was used to increase the second hyperpolarizability\cite{may07.01}.

\begin{figure}\center
\includegraphics[width=4in]{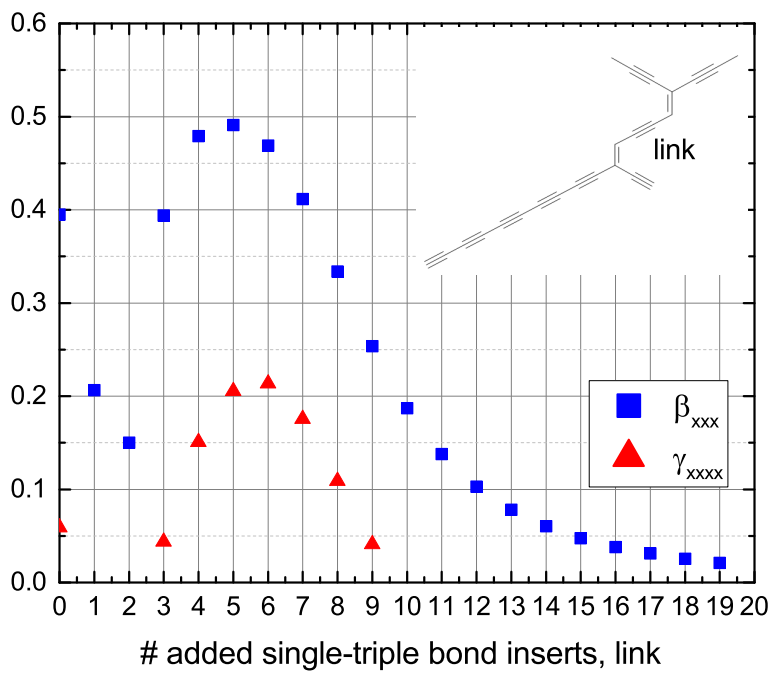}
\caption{Variation of $\beta_{xxx}$ and $\gamma_{xxxx}$ for a donor-acceptor structure with a long end group and a varying link edge.}\label{fig:da_2star_linker}
\end{figure}

Figure \ref{fig:da_2star_linker} shows the variation in $\beta_{xxx}$ and $\gamma_{xxxx}$ as the link between the two bent wires (double bonds) is extended.  If we assign values typical of single, double, and triple Carbon bonds, then the lengths of the fixed edges of the graph in the inset figure are calculable, as is the variable length of the linker as additional single-triple bond extensions are added.  The result is a plot showing the existence of an ideal linker for $\beta_{xxx}$ (and a slightly longer one for $\gamma_{xxxx}$).  These results are again suggestive of a design rule for donor-acceptor molecules, whereby a short side group is placed some distance from the donor or acceptor end of the main chain, and a linker of about equal length is placed between that side group and another located near the other end of the chain.  With this configuration, the shape of the eigenstates are distorted as displayed in the upper right panel of Figure \ref{fig:1prong_states_quadPlot}, generating a large change in dipole moment while maintaining a large overlap among the ground state and the lowest eigenstates.

It is now clear that the graph model suggests that a second design rule for nonlinear optical structures is to place a defect or side group on a wire somewhere between an end and the middle, but not at the end of the chain, as is common practice for creating donor-acceptor molecules.  Moreover, the presence of potential-altering groups along a chain will do nothing to enhance the optical nonlinearities, but it can convert a superscaling spectrum into a non-optimum scaling spectrum and reducing the nonlinearities by many orders of magnitude.  Conventional molecular design rules may require some rethinking, but the exact details of how the design rules should be stated awaits a many-electron calculation.

\section{Dressed graphs with side groups}\label{sec:deltaANDprong}

\begin{figure}\center
\includegraphics[width=2.5in]{Wire_bare.eps}\includegraphics[width=2.5in]{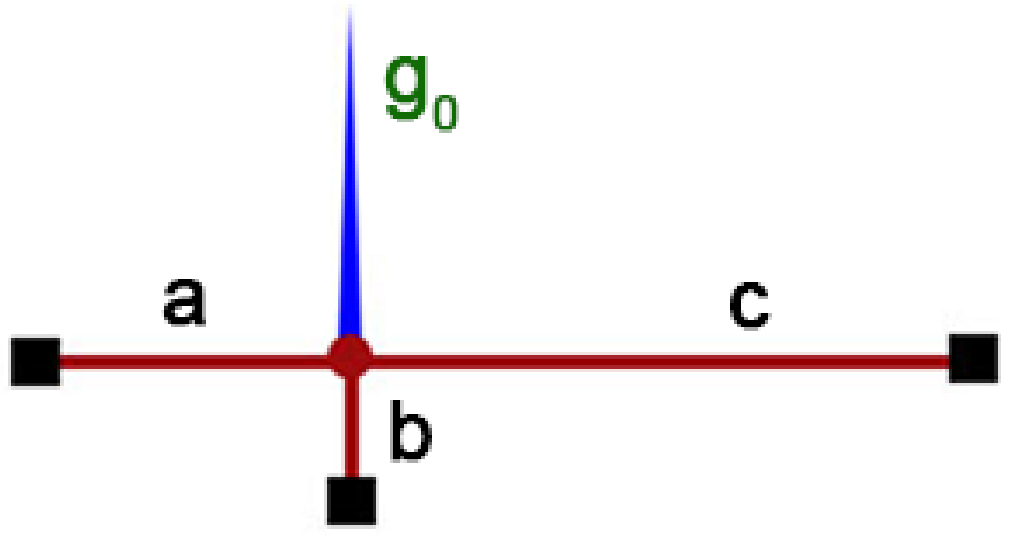}\\
\includegraphics[width=2.5in]{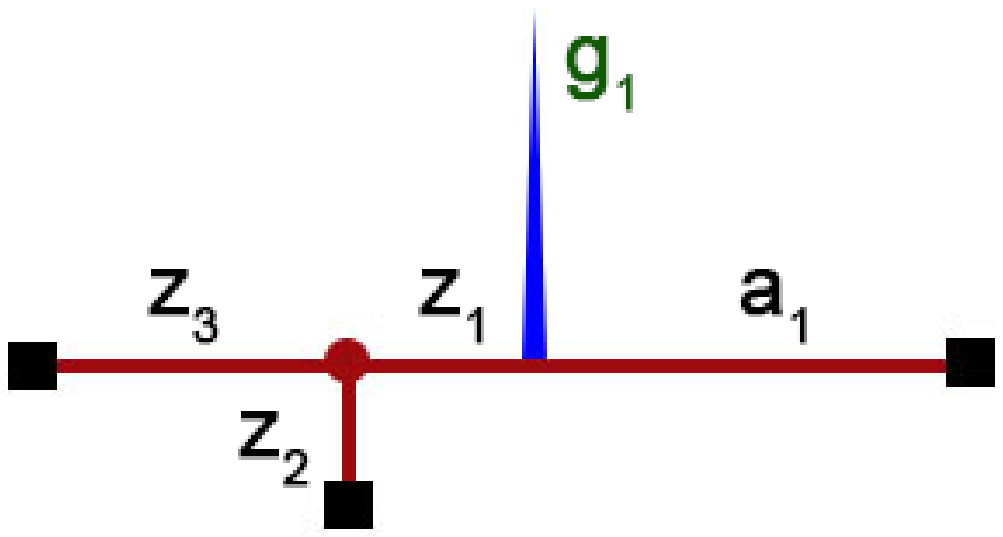}\includegraphics[width=2.5in]{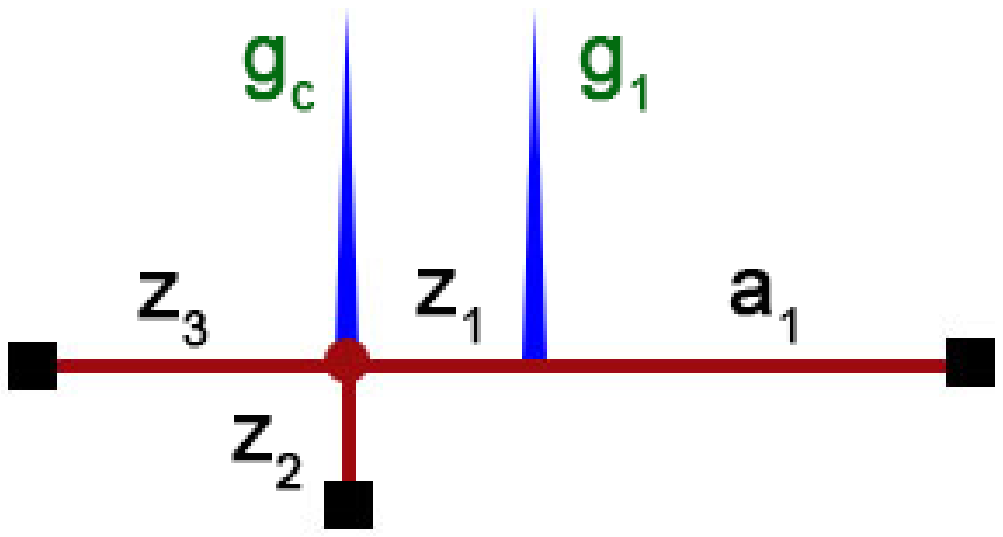}\\
\caption{The bare wire graph and three variants combining prongs and $\delta$ potentials.}
\label{fig:DeltaProngWireGraphs}
\end{figure}

The presence of side groups and defects on a wire, at the same time is important to investigate, as it represents a realistic outcome for the design of a quantum wire or quasi-linear molecule.  Figure \ref{fig:DeltaProngWireGraphs} shows a bare wire and three variants, each with both $\delta$ functions and a prong attached.

We examine each in turn.

\subsection{One $\delta$ at one prong}\label{1deltaAT1prongWire}

The upper right panel of Figure \ref{fig:DeltaProngWireGraphs} shows a one prong wire to which a finite potential of the form $V(x)=(g/L)\delta (x-a)$ has been placed at the central vertex of a star.  The solution to this graph has been published \cite{lytel13.01}.  The edge functions match at the central vertex, but the presence of the delta function changes the flux conservation condition \cite{kuchm04.01,kostr99.01,exner89.01} from its canonical form \cite{pasto09.01,lytel13.01} to

\begin{equation}\label{3starDeltaSecular}
F_{star}(a,b,c) +(2g/kL)\sin{ka}\sin{kb}\sin{kc}=0 .
\end{equation}

The solutions to Eq. (\ref{3starDeltaSecular}) for positive energies (real k) may be found numerically and are described in detail in ref. \cite{lytel13.01}.

\subsection{One $\delta$ on one prong}\label{1deltaAND1prongWire}

The lower left panel of Figure \ref{fig:DeltaProngWireGraphs} shows a one prong wire to which a finite potential of the form $V(x)=(g/L)\delta (x-a)$ has been placed on one prong.  This is topologically a star graph with a $\delta$ on one arm. For the 3-star with edges $z_{1},z_{2},z_{3}$, the secular function $F_{star}(z_{1},z_{2},z_{3})$ for the graph is written in a form that is easy to use with the $\delta$ motif to make bigger graphs.  This is
\begin{eqnarray}\label{3starF}
F_{star}(z_{1},z_{2},z_{3}) &=& \sin{kz_{1}}\sin{kz_{2}}\cos{kz_{3}}\nonumber \\
&+& \sin{kz_{1}}\cos{kz_{2}}\sin{kz_{3}} \\
&+& \cos{kz_{1}}\sin{kz_{2}}\sin{kz_{3}}\nonumber
\end{eqnarray}
For a single unterminated end A, conservation of flux at the central vertex $Z$ produces the following secular equation relating the amplitudes at the end A and the central amplitude Z:
\begin{equation}\label{3starSec1Unterm}
ZF_{star}(z_{1},z_{2},z_{3}) = A\sin kz_{2}\sin kz_{3}
\end{equation}
The right-hand side is the net flux through its unterminated vertex A required to conserve flux at the central vertex.  For the $\delta$ motif, with one end attached to the star at vertex Z and the other end terminated, we have\cite{lytel13.04}
\begin{equation}\label{1deltaF}
F_{\delta}(g_{1};a_{1},z_{1}) = \frac{2g_{1}}{kL}\sin{ka_{1}}\sin{kz_{1}}+\sin{k(a_{1}+z_{1})}.
\end{equation}
where we have normalized the $\delta$ potential to the \emph{total} length $L=a_{1}+z_{1}+z_{2}+z_{3}$ of the entire graph, as previously discussed.

We now combine the star motif with the $\delta$ motif to get the pair of equations for the amplitudes A and Z:
\begin{eqnarray}\label{1deltaStarAmpEqns}
AF_{\delta_{1}}(g_{1};a_{1},z_{1}) &=& Z\sin{ka_{1}} \\
ZF_{star}(z_{1},z_{2},z_{3}) &=& A\sin{kz_{2}}\sin{kz_{3}}\nonumber
\end{eqnarray}
Setting the determinant of Eqs. (\ref{1deltaStarAmpEqns}) to zero, we get the secular function for this composite graph:
\begin{eqnarray}\label{1starDeltaSecular}
F_{star-\delta_{1}}&=&F_{star}(z_{1},z_{2},z_{3})F_{\delta_{1}}(g_{1};a_{1}z_{1})\nonumber \\
&-& \sin{ka_{1}}\sin{kz_{2}}\sin{kz_{3}}
\end{eqnarray}

We will rewrite this in a form that allows us to extract a common factor $\sin{kz_{1}}$ and remove it from the characteristic equation.  This factor is spurious, as it disconnects the graph.  Using the expressions for the star motif, we get
\begin{eqnarray}\label{1starDeltaF}
F_{star-\delta_{1}}&=&F_{star}(z_{1},z_{2},z_{3})\frac{2g_{1}}{kL_{tot}}\sin{ka_{1}}\sin{kz_{1}}\nonumber \\
&+&F_{star}(z_{1},z_{2},z_{3})\sin{k(a_{1}+z_{1})}\nonumber \\
&-&\sin{ka_{1}}\sin{kz_{2}}\sin{kz_{3}}
\end{eqnarray}
where $L_{tot}=a_{1}+z_{1}+z_{2}+z_{3}$ is the total length of the graph and is unchanged, regardless of the position of the $\delta$ function on the one prong.

Using Eq. (\ref{3starF}), the second and third lines in Eq. (\ref{1starDeltaF}) may be combined as
\begin{eqnarray}\label{1starDeltaF2}
&& F_{star}(z_{1},z_{2},z_{3})\sin{k(a_{1}+z_{1})} \\
&-&\sin{ka_{1}}\sin{kz_{2}}\sin{kz_{3}}\nonumber \\
&=& \sin{kz_{1}}\sin{k(z_{2}+z_{3})}\sin{k(a_{1}+z_{1})}\nonumber \\
&+& \cos{kz_{1}}\sin{kz_{2}}\sin{kz_{3}}\sin{k(a_{1}+z_{1})}\nonumber \\
&-& \sin{ka_{1}}\sin{kz_{2}}\sin{kz_{3}}\nonumber \\
&=& \sin{kz_{1}}\sin{k(a_{1}+z_{1})}\sin{kz_{2}}\cos{kz_{3}}\nonumber \\
&+& \sin{kz_{1}}\sin{k(a_{1}+z_{1})}\cos{kz_{2}}\sin{kz_{3}}\nonumber \\
&+& \sin{kz_{2}}\sin{kz_{3}}\left[\cos{kz_{1}}\sin{k(a_{1}+z_{1})}-\sin{ka_{1}}\right]\nonumber \\
&=& \sin{kz_{1}}\sin{k(a_{1}+z_{1})}\sin{kz_{2}}\cos{kz_{3}}\nonumber \\
&+& \sin{kz_{1}}\sin{k(a_{1}+z_{1})}\cos{kz_{2}}\sin{kz_{3}}\nonumber \\
&+& \sin{kz_{2}}\sin{kz_{3}}(\cos^{2}{kz_{1}}-1)\sin{ka_{1}}+\sin{kz_{1}}\cos{ka_{1}}\cos{kz_{1}}\nonumber \\
&=& \sin{kz_{1}}\sin{k(a_{1}+z_{1})}\sin{kz_{2}}\cos{kz_{3}}\nonumber \\
&+& \sin{kz_{1}}\sin{k(a_{1}+z_{1})}\cos{kz_{2}}\sin{kz_{3}}\nonumber \\
&+& \sin{kz_{1}}\sin{kz_{2}}\sin{kz_{3}}\cos{k(a_{1}+z_{1})}\nonumber \\
&=& \sin{kz_{1}}F_{star}(a_{1}+z_{1},z_{2},z_{3})\nonumber
\end{eqnarray}

Combining the last two equations, we get a nice compact form for the secular function for the star graph with a $\delta$ function in one arm:
\begin{eqnarray}\label{1starDeltaFfinal}
&& F_{star-\delta_{1}}(g_{1};a_{1},z_{1},z_{2},z_{3})=\sin{kz_{1}}\nonumber \\
&\times & \left[F_{star}(z_{1},z_{2},z_{3})\frac{2g_{1}}{kL_{tot}}\sin{ka_{1}}+F_{star}(a_{1}+z_{1},z_{2},z_{3})\right]\nonumber
\end{eqnarray}
which is how it must be, because when $g=0$, the $\delta$ function is absent, and the graph becomes a 3-star but with the first prong of length $a_{1}+z_{1}$.

The nonlinearities of this graph, calculated from a Monte Carlo computation, have the extreme values shown in Figure \ref{fig:resultsTable}.

\subsection{Two $\delta$s, one at a prong}\label{2deltaOneAtAprong}

The lower right panel of Figure \ref{fig:DeltaProngWireGraphs} shows a star with a $\delta$ at its center and one on its arm.

As we've seen, for the 3-star with prongs $(a,b,c)$ dressed at the center with a $\delta$ function of strength $g_{c}/L$ with $L=a+b+c$, the secular function becomes \cite{lytel13.01}
\begin{equation}\label{3star1delta1C}
F_{star\delta_{c}}=F_{star}(a,b,c) +(2g_{c}/kL)\sin{ka}\sin{kb}\sin{kc}.
\end{equation}

The motif method immediately yield the secular function for the star with a $\delta$ at the center and one on one prong by using $F_{star\delta_{c}}$ instead of $F_{star}$ in the amplitude equations in Eq. (\ref{1deltaStarAmpEqns}).  The result is

\begin{eqnarray}\label{star2delta1P1C}
F_{star-\delta_{c}\delta_{1}}&=& F_{star\delta_{c}}(g_{c},z_{1},z_{2},z_{3})F_{\delta_{1}}(g_{1};a_{1},z_{1})\nonumber \\
&-&\sin{ka_{1}}\sin{kz_{2}}\sin{kz_{3}}
\end{eqnarray}

Finally, we may write this in a simplified form by using Eq. (\ref{3star1delta1C}.  Factoring out a common factor $\sin{kz_{1}}$, we get

\begin{equation}\label{star2delta1P1Cexpand}
F_{star-\delta_{c}\delta_{1}}(g_{1},g_{c};z_{1},z_{2},z_{3},a_{1})= \sin{kz_{1}} \times G;\nonumber \\
\end{equation}
where

\begin{eqnarray}\label{G}
&&G_{star-\delta_{c}\delta_{1}}(g_{1},g_{c};z_{1},z_{2},z_{3},a_{1})\nonumber \\
&=& F_{star}(z_{1},z_{2},z_{3})\frac{2g_{1}}{kL_{tot}}\sin{ka_{1}}\nonumber \\
&+& F_{star}(a_{1}+z_{1},z_{2},z_{3}) \\
&+& \frac{2g_{c}}{kL_{tot}}\sin{k(a_{1}+z_{1})}\sin{kz_{2}}\sin{kz_{3}}\nonumber \\
&+&\frac{4g_{1}g_{c}}{{kL}^{2}}\sin{ka_{1}}\sin{kz_{1}}\sin{kz_{2}}\sin{kz_{3}}\nonumber
\end{eqnarray}
where $L_{tot}=a_{1}+z_{1}+z_{2}+z_{3}$ is the total length of the graph.

\section{Conclusions}\label{sec:end}

\begin{figure}\center
\includegraphics[width=5in]{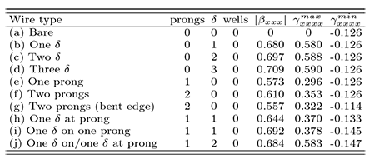}
\caption{Maximum $|\beta_{xxx}|$, and maximum and minimum $\gamma_{xxxx}$ for the wire graphs in Figs \ref{fig:DeltaWireGraphs}, \ref{fig:ProngWireGraphs}, and \ref{fig:DeltaProngWireGraphs}.}\label{fig:resultsTable}
\end{figure}

All of the variants of the bare wire graph shown in Figures \ref{fig:DeltaWireGraphs}, \ref{fig:ProngWireGraphs}, and \ref{fig:DeltaProngWireGraphs} have favorable geometries (configurations of the edges) to produce very large first and second hyperpolarizabilities, as shown in Figure \ref{fig:resultsTable}.  All have topologies that generate random spectra but with fixed gaps set by a quadratic scaling law.  For each class of graphs, there exist ideal geometries such that the shape of the eigenfunctions may be tailored to provide a giant enhancement in their nonlinear optical responses.

Structures with large intrinsic second and third hyperpolarizabilities necessarily have superscaling spectra whose energy $E_{n}$ scales with mode number n as $E\sim n^{\mu}$, with $\mu\geq 2$.  This result arises solely from quantum physics and holds for any system whose nonlinearities may be described by a sum over states.  This follows from extensive Monte Carlo analysis of a very large space of spectra and moments constrained only by the Thomas-Reich-Kuhn sum rules, which themselves are consequences of the validity of the completeness of the eigenfunctions, ie, of the self-adjointness of the underlying Hamiltonian, and the validity of the canonical commutation relations.  It is thus a necessary and fundamental design rule that \emph{highly active nonlinear optical molecules must have superscaling spectra}.  In a nutshell, this conclusion explains at once the giant gap between the fundamental limits and all of the measured optical nonlinearities to date.  But superscaling spectra are not sufficient for large optical nonlinearities.

In this paper, we have shown by use of quantum graph models that systems designed with electron flow pathways away from their main chains will create \emph{sufficient spatial separation} among the ground state and first few excited states to generate large differences $\bar{x}_{22}$ and $\bar{x}_{11}$  in their dipole moments required for the nonlinear optical response to approach the fundamental limits, \emph{provided that the geometry of the structure is optimized}.  In fact, quantum loop graphs have superscaling spectra but very small nonlinearities \cite{shafe12.01} precisely because their geometry is almost maximally non-optimum.  But linear and quasi-linear graphs, resembling the geometries of conjugated chains with side groups, have near-optimum geometries, and when the flux along the main chains is interrupted (ie, discontinuous across the chain), the ground and first few excited states are spatially separated, creating a large shift in the flow of charge upon a virtual transition to an excited state from the ground state, and generating a large nonlinear optical response.

The two design rules (1) superscaling spectra and (2) spatially separated ground and excited states that still maintain good spatial overlap, are heuristics for designing new molecules and nanowires with large hyperpolarizabilities.  They do not explain \emph{how} to design such molecules, but they do explain why most prior designs fail to achieve large values.  The heuristics also point toward starting points for designing new molecules.  For example, structures containing many electrons filling single particle states to a Fermi level, will have large responses if the transition moments of the electron at the top of the Fermi level to unfilled states above it dominate all other transitions from electrons below the Fermi level to states above it.  In this case, the system behaves as a quasi-one-electron structure, and the results in this paper correspond to physically viable models.  Such systems, designed as networks of structures emulating a quantum graph with a quasi-linear geometry and a side prong, will have both superscaling spectra and the transition moments required for large hyperpolarizabilities.

\section*{Acknowledgments}
SMM and MGK thank the National Science Foundation (ECCS-1128076) for generously supporting this work.

\appendix
\setcounter{figure}{0}
\renewcommand{\thefigure}{A\arabic{figure}}
%
\section{Off resonance square wells and steps}\label{sec:squareWells}

\begin{figure}\center
\includegraphics[width=3in]{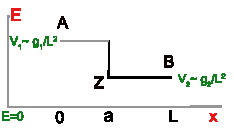}\\
\caption{Motif for the step on a wire.}\label{fig:motifWell}
\end{figure}

The graphs in the main body of the text were wires with zero potential (bare graphs) or with $\delta$ function potentials (dressed graphs).  Since the $\delta$ function operates at a single point along the wire, its spectrum and that of bare graphs scale with energy eigenfunction number in similar ways. For small, positive $\delta$ strength $g$ the two topologies are nearly identical.  The effect in each case is to produce a spatial shift in the ground and first few excited states whose magnitude depends either on the length of the prong or the strength g of the $\delta$ function. This appendix shows how wires dressed with potentials of finite spatial extent can produce the same spatial shifts required to generate large nonlinearities.  Dressed graphs with a finite potential $V(x)$ are wells or steps.  We consider here the square ($V(x)=constant$) and slant ($V(x)\sim x$ well potentials.  The analysis shows behavior similar to those of the bare and $\delta$ dressed wires, with hyperpolarizabilities capable of approaching the fundamental limits.  This is not surprising, since the dominant feature creating the required spatial separation among the lowest eigenfunctions is anything that causes a spatial kink in the shape of the eigenfunction at an appropriate location on the wire.
Figure \ref{fig:motifWell}, left, shows the motif graph for the square well.  Flux enters and exits at the ends and is conserved at the center vertex Z.  The edge states are written in canonical form that matches flux at the vertices:
\begin{eqnarray}\label{edgeSquareMotif}
\phi_{L}(x)&=&\frac{A\sin{k_{1}(a-x)}+Z\sin{k_{1}x}}{\sin{k_{1}a}} \\
\phi_{R}(x)&=&\frac{B\sin{k_{2}(x-a)}+Z\sin{k_{2}(L-x)}}{\sin{k_{2}(L-a)}}\nonumber
\end{eqnarray}
Conservation of flux yields
\begin{equation}\label{fluxCon}
ZF_{sq}(k_{1},k_{2};a,b)=Ak_{1}\sin{k_{2}b}+Bk_{2}\sin{k_{1}a}
\end{equation}
with a secular function
\begin{equation}\label{secSqWell}
F_{sq}(k_{1},k_{2};a,b)=k_{1}\cos{k_{1}a}\sin{k_{2}b}+k_{2}\cos{k_{2}b}\sin{k_{1}a}
\end{equation}
and we have set $b=L-a$.  The wave numbers are $k_{i}=\sqrt{2m(E-V_{i})/\hbar^{2}}$ and $V_{i}=(\hbar^{2}/m)g_{i}/L^2$ for $i=1,2$.

\begin{figure}\center
\includegraphics[width=2.5in]{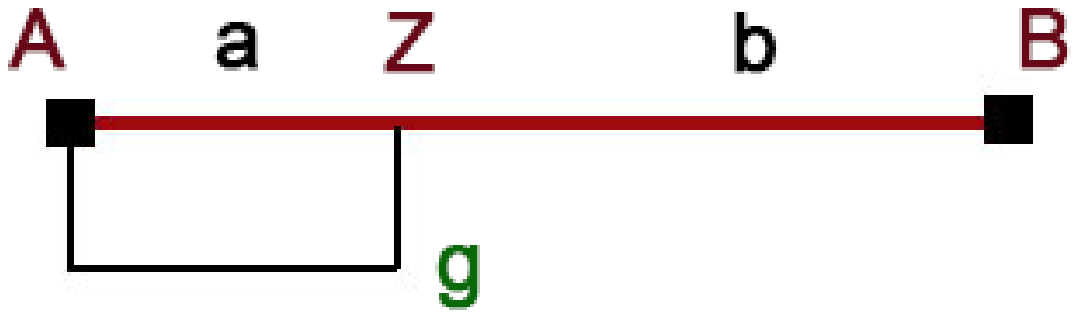}\includegraphics[width=2.5in]{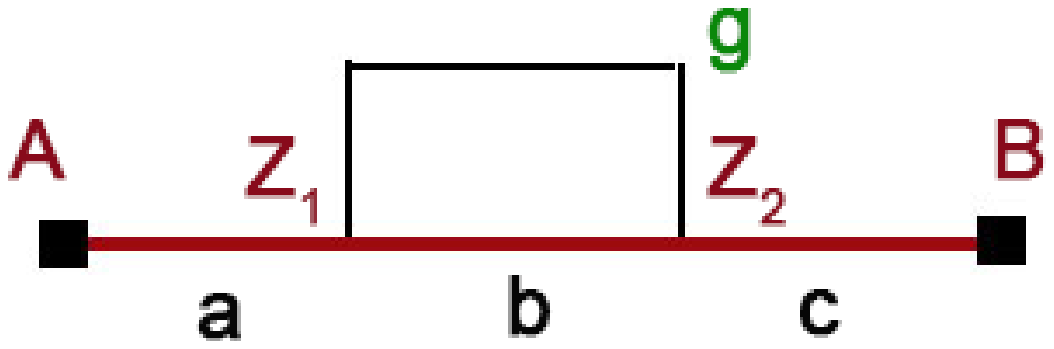}\\
\caption{Graph for a square well on a wire (left) and a square step on a wire (right).}\label{fig:WireWellGraphs}
\end{figure}

\begin{figure}\center
\includegraphics[width=4in]{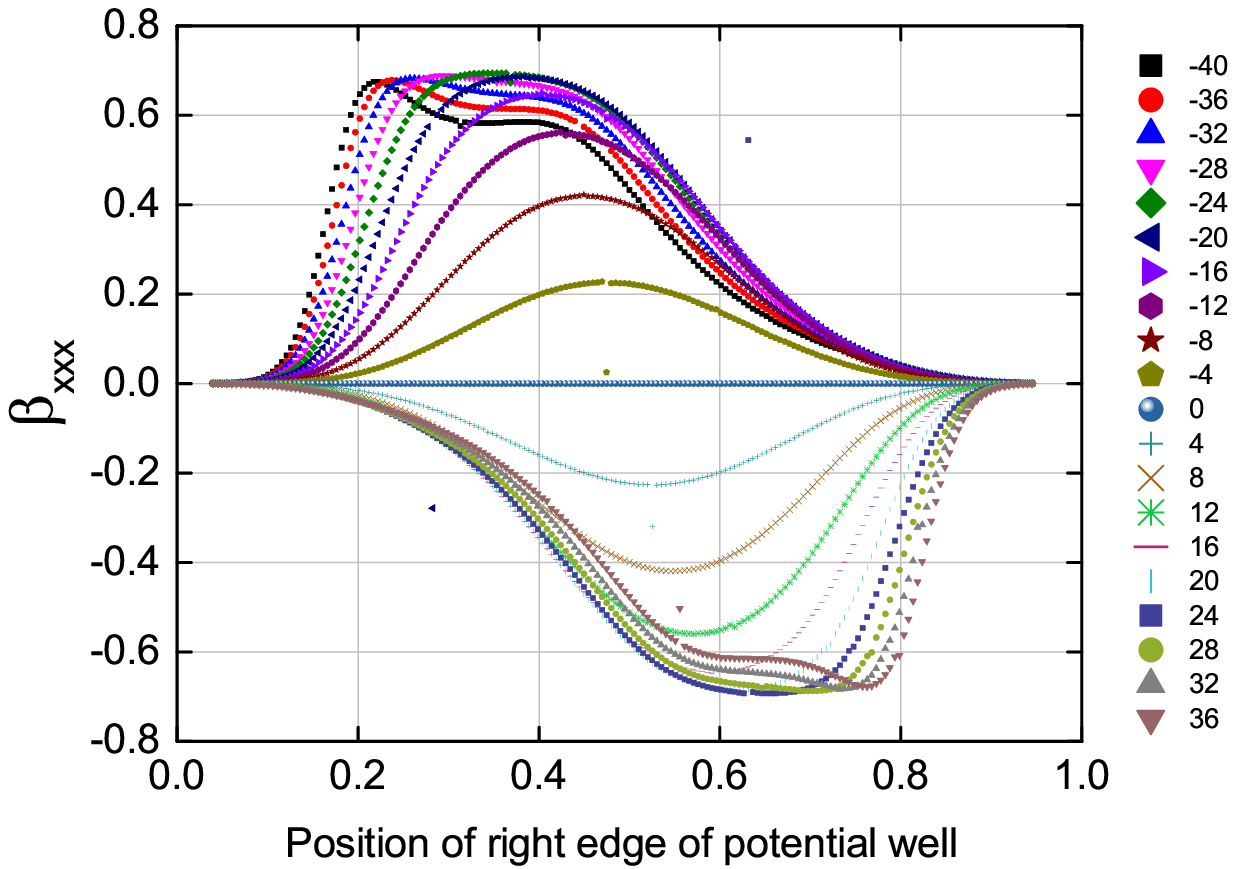}
\caption{Variation of the first hyperpolarizability $\beta_{xxx}$ with the position of the right edge of the square step ($g>0$) or well ($g<0)$ potential for twenty values of the potential strength g shown in the legend.}\label{fig:1well20curvesBetaVSaATgZero=-40to40}
\end{figure}

\begin{figure}\center
\includegraphics[width=4in]{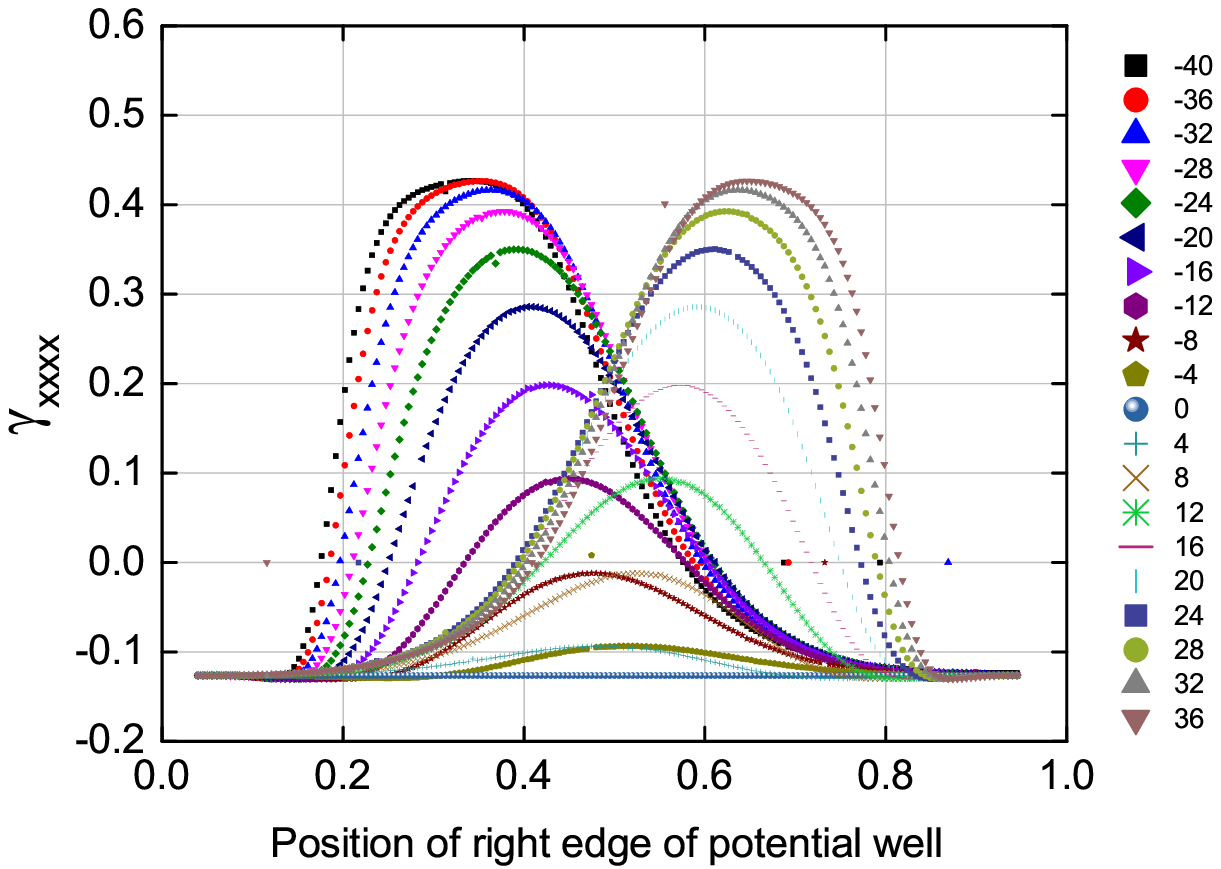}
\caption{Variation of the first hyperpolarizability $\beta_{xxx}$ with the position of the right edge of the square step ($g>0$) or well ($g<0)$ potential, again for twenty values of g shown in the legend.}\label{fig:1well20curvesGammaVSaATgZero=-40to40}
\end{figure}

Consider the square well in Fig \ref{fig:WireWellGraphs}, left.  The potential for $0\leq x\leq a$ is $V(x)=V_{0}=(\hbar^{2}/m)g/L^{2}$ and zero elsewhere.  Note that positive $g$ corresponds to a barrier,and all energies are positive.  Negative $g$ corresponds to a well, where negative energies are possible.  Define the wavenumbers for the nonzero potential region as $K^2=(k^{2}-\alpha^{2})$, where $\alpha=\sqrt{2g/L^{2}}$.  The edge functions are
\begin{eqnarray}\label{edges}
\phi_{L}(x)&=&A\sin{Kx},\ 0\leq x\leq a\\
\phi_{R}(x)&=&B\sin{k(L-x}),\ a\leq x\leq L\nonumber
\end{eqnarray}
The secular equation is found by matching the wavefunction and its derivative at the potential boundary.  The result is
\begin{equation}\label{secular}
k\cos{k(L-a)}\sin{Ka}+K\sin{k(L-a)}\cos{Ka}=0
\end{equation}
It may be seen that this is identical to that obtained by using Eqs. (\ref{fluxCon}) and Eq. (\ref{secSqWell}), with $A=B=0$.  This may be solved numerically to yield the spectra.  The edge functions are then used to compute the transition moments.  The results are displayed in Figures \ref{fig:1well20curvesBetaVSaATgZero=-40to40} for $\beta_{xxx}$ and \ref{fig:1well20curvesGammaVSaATgZero=-40to40} $\gamma_{xxx}$.  They are strikingly similar to those obtained for a $\delta$ wire with $g<0$, as advertised.

For completeness, we consider the step well in Figure \ref{fig:WireWellGraphs}, right.  The flux conservation equations are easily found using the motif in Eq. (\ref{secSqWell}).  They are
\begin{eqnarray}\label{fluxConStep}
Z_{1}F_{sq}(k_{1},k_{2};a,b)&=&Z_{2}k_{2}\sin{k_{1}a} \\
Z_{2}F_{sq}(k_{1},k_{2};c,b)&=&Z_{1}k_{2}\sin{k_{1}c}\nonumber
\end{eqnarray}
where the wavenumbers on either side of the well are $k_{1}=\sqrt{2mE/\hbar^2}$ and the center wavenumber is $k_{2}=\sqrt{2m(E-V_{0})/\hbar^2}$.  The secular equation becomes
\begin{eqnarray}\label{secSqStep}
&&F_{step}(k_{1},k_{2};a,c,b) \\
&=&F_{sq}(k_{1},k_{2};a,b)F_{sq}(k_{1},k_{2};c,b)\nonumber \\
&-&k_{2}^{2}\sin{k_{1}a}\sin{k_{1}c}=0\nonumber
\end{eqnarray}
After a little algebra, a factor $\sin{k_{2}b}$ may be removed.  Writing $F_{step}=\sin{k_{2}b}\times G_{step}$, we get
\begin{eqnarray}\label{secSqStep2}
G_{step}&=&\left[k_{1}^{2}\cos{k_{1}a}\cos{k_{1}c}-k_{2}^{2}\sin{k_{1}a}\sin{k_{1}c}\right]\sin{k_{2}b}\nonumber \\
&+&k_{1}k_{2}\sin{(a+c)k_{1}}\cos{k_{2}b}
\end{eqnarray}
The spectra may be obtained by solving $G_{step}=0$ for the energies.  The results are similar to those for the square well and are not presented here.
%
%
%
\setcounter{figure}{0}
\renewcommand{\thefigure}{B\arabic{figure}}

\section{Off-resonance slant well}\label{sec:slantWells}

\begin{figure}\center
\includegraphics[width=3.4in]{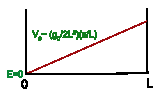}
\caption{Slant well potential.}\label{fig:slant_well}
\end{figure}

Consider a slant well with infinite barriers at $x=0$ and $x=L$, as shown in Figure \ref{fig:slant_well}.  The potential energy (in units where $\hbar=m=e=1$ is $V_{0}=(g_{0}/2L^3)x$, where $g_{0}$ is dimensionless.
The wave equation takes the form
\begin{equation}\label{slantwellWaveEqn}
-\frac{1}{2}d^2\psi(x)/dx^2+\left[V(x)-E\right]\psi(x)=0
\end{equation}
Define the energy as $E=k^{2}/2$ and a shifted variable $\xi$ as
\begin{equation}\label{xShift}
\xi=\left(\frac{g_{0}}{L^3}\right)^{1/3}\left[x-\frac{(kL)^2}{g_{0}}L\right]
\end{equation}
Then the wave equation becomes the Airy equation $\psi''(\xi)-\xi\psi(\xi)=0$.  The solutions are
\begin{equation}\label{states}
\psi(x)=aAi(\xi)+bBi(\xi),
\end{equation}
where $\xi(x)$ is given in Eq. (\ref{xShift}) and a and b are constants to be determined by the boundary conditions that the wave function vanish at $x=0$ and $x=L$:
\begin{eqnarray}\label{ampEqns}
aAi(\xi_{0})+bBi(\xi_{0})=0 \\
aAi(\xi_{L})+bBi(\xi_{L})=0\nonumber
\end{eqnarray}
where
\begin{eqnarray}\label{endpts}
\xi_{0}=-(kL)^2/g_{0}^{2/3} \\
\xi_{L}=g_{0}^{1/3}\left[1-(kL)^2/g_{0}\right] \nonumber
\end{eqnarray}

From Eq. (\ref{ampEqns}), we get the secular equation that determines the eigenvalues $k_{n}$:
\begin{equation}\label{secEqn}
Ai(\xi_{0})Bi(\xi_{L})-Ai(\xi_{L})Bi(\xi_{0})=0,
\end{equation}
The solutions to Eq. (\ref{secEqn}) determine the eigenvalues of the slant well.

\begin{figure}\center
\includegraphics[width=4.0in]{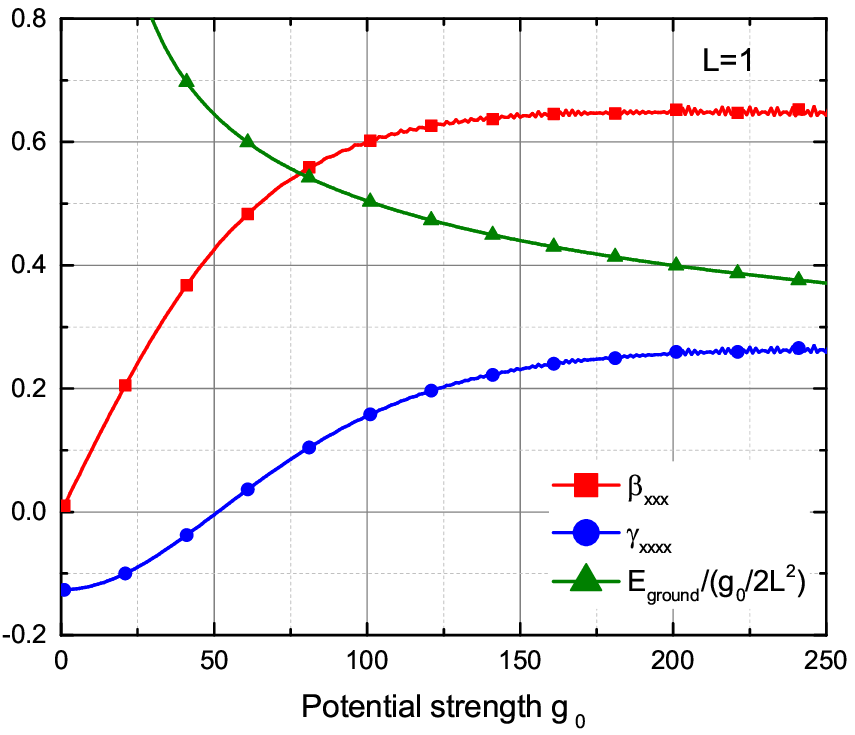}
\caption{Dependence of $\beta_{xxx}$ and $\gamma_{xxxx}$ on the potential strength $g_{0}$ of a slant potential well.}\label{fig:slantWell_results}
\end{figure}

Fig \ref{fig:slantWell_results} plots the hyperpolarizabilities against the potential strength $g_{0}$, as well as the ratio of the ground state energy to the potential energy at $g_{0}$.  Both the first and second hyperpolarizabilities saturate as the well gets higher.

The eigenfunctions for graphs with $g_0=0$, $g_0=50$, and $g_0=200$ are shown below.  The effect of a slant well potential is to break the centrosymmetry of the $g_{0}=0$ graph, localizing the ground state of the electron to the left of the well below the barrier energy, while pushing the excited states to the right, creating a large transition moment and leading to large responses.

\begin{figure}\center
\includegraphics[width=4in]{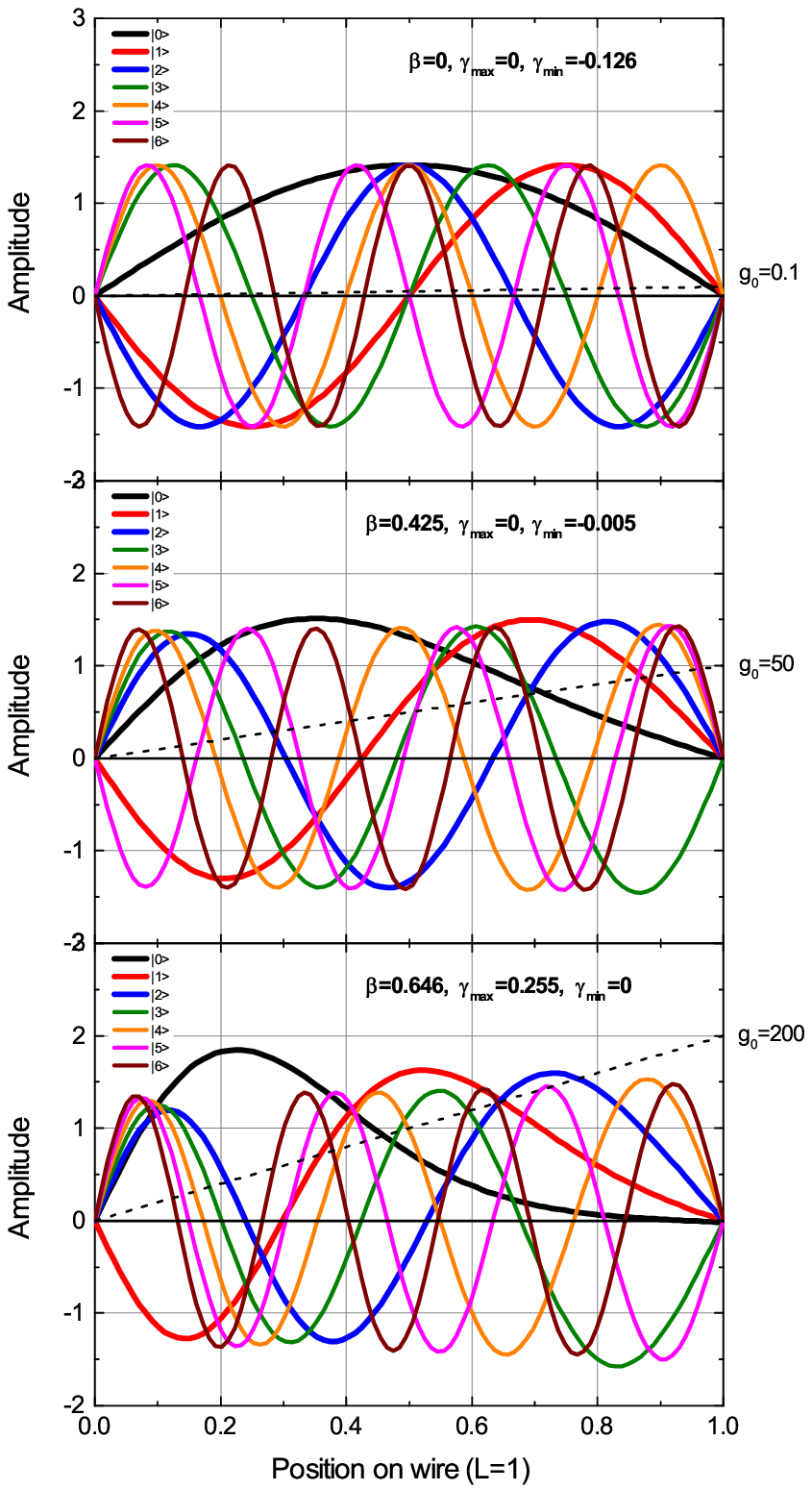}
\caption{First seven eigenfunctions of the slant well at zero potential, ie, a bare wire.  The graph is centrosymmetric with zero $\beta_{xxx}$ and the well known value $\gamma_{xxxx}=-0.126$.}\label{fig:slantWell_eigenfunctions_gZero=0.1_50_200_vertical}
\end{figure}

The problem is easily extended to negative potentials, $g_{0}<0$.  The shifted position variable becomes
\begin{equation}\label{xShiftneg}
\xi=-\left(\frac{|g_{0}|}{L^3}\right)^{1/3}\left[x+\frac{(kL)^2}{|g_{0}|}L\right]
\end{equation}
and the endpoint values become
\begin{eqnarray}\label{endptsneg}
\xi_{0}=-(kL)^2/|g_{0}|^{2/3} \\
\xi_{L}=-|g_{0}|^{1/3}\left[1+(kL)^2/|g_{0}|\right] \nonumber
\end{eqnarray}

The secular equation remains unchanged, except for the replacement of the endpoint values in Eq. (\ref{endpts}) with those in Eq. (\ref{endptsneg}).  There are negative energy solutions for sufficiently large $g_{0}$ that are easily discovered by solving the secular equation, with $k\rightarrow \imath k$.

\end{document}